# Imaging of Four Galactic Supernovae Remnants in the Mid-Infrared, and their Interaction with the Interstellar Medium

R.A. Marquez-Lugo, J.P. Phillips

Instituto de Astronomía y Meteorología, Av. Vallarta No. 2602, Col. Arcos Vallarta, C.P. 44130 Guadalajara, Jalisco, México   e-mails : alejmar.astro@gmail.com, jpp@astro.iam.udg.mx

**Abstract**

We provide mid-infrared (MIR) imaging, photometry and profiles for the Galactic supernova remnants (SNRs) G001.0-00.1, G355.9-02.5, G355.6-00.0, and W28 based upon data deriving for the Galactic Legacy Infrared Midplane Survey Extraordinaire (GLIMPSE). All of the sources show evidence for interaction with the interstellar medium (ISM), leading to curved frontal structures and apparent voids in the ISM. An analysis of the spectral energy distributions (SEDs) of Class I young stellar objects (YSOs) within the north-westerly interaction region of W28, and of the density of stars within the borders of the SNR, suggests that many of them may have been triggered by the SN event. 2-D radiative transfer modelling permits us to constrain the physical parameters of the sources. It is also noted that the location of Class I YSOs about the perimeter of G001.0-00.1, and close to frontal arcs associated with SNR G355.9-02.5, suggests that star formation may have been triggered by these SNRs as well. Finally, it is found that the MIR colours of the frontal structures appear consistent with shock excitation of the v = 0→0 transitions of $H_2$, although it is conceivable that emission by polycyclic aromatic hydrocarbons (PAHs) may also play a role. Where the latter mechanism is relevant, then it is possible that emission derives from the shattering of larger grains in frontal regions, leading to increased volume densities of PAH carrying grains.

**Key Words:** stars: formation --- (stars:) supernovae: general --- ISM: kinematics and dynamics --- ISM: jets and outflows --- ISM: clouds --- (ISM:) supernova remnants

## 1. Introduction

The expansion of supernovae remnants (SNRs) into the interstellar medium (ISM) leads to important consequences for the ISM, for star formation within the Galaxy, and for the structures and kinematics of the SNRs themselves. Thus, interaction between the SNRs and interstellar gas results in a strong braking of shell expansion where diameters exceed ≈ 20 pc (the snow-plough process; see e.g. Van der Swaluw 1971), and distortion of the envelopes where they meet inhomogeneities within the ISM; the latter resulting from compression by cooler and denser regions of ambient gas, or accelerated expansion ("blow-outs") where the ISM is hotter and less dense (see e.g. Caswell 1977, Caswell et al. 1983; and Gray 1994a, and the further discussion of SNRs below). Similarly, IS clouds are shredded and destroyed where shell velocities exceed 50 km s$^{-1}$ (e.g. Boss 1995; Foster & Boss 1996, 1997; Vanhala & Cameron 1998; Fukuda & Hanawa 2000), a circumstance which may occur within 10 and 100 pc of the supernova event, depending upon the evolutionary state of the pre-impact cores (e.g. Oey & García-Segura 2004; Vanhala & Cameron 1998). Lower velocities at larger distances, by contrast, are prone to lead to compression of the clouds, gravitational instability, and the triggering of local star formation (e.g. Boss 1995; Foster & Boss 1996, 1997; Nakamura et al. 2006; Melioli et al. 2006; Boss et al. 2008).

There has, up to the present, been relatively little evidence for infrared emission deriving from SNRs. Previous surveys with IRAS, for instance, have lead to the detection of only ~20 % of sources (Arendt 1989; Saken et al. 1992), whilst similarly low proportions of SNR have been detected in the mid-infrared (MIR) (Reach et al. 2006). Nevertheless, it is clear that a variety of mechanisms may lead to emission at these wavelengths. Reach et al. (2006) point out that shock interaction with interstellar (IS) gas leads to cooling in a variety of fine structure lines, whilst shock excited $H_2$ and ionic transitions have been observed in several SNR (see e.g. Arendt et al. 1999; Oliva et al. 1999; Burton et al. 1988; Richter, Graham & Wright 1995; Rosado et al. 2007). Shattering of grains within strong shocks may also lead to the formation of higher densities of smaller grains (Allain, Leach & Sedlmayr 1996; Jones et al. 1996), responsible for polycyclic aromatic hydrocarbon (PAH) emission bands at 3.3, 6.2, 7.7 ad 8.6 μm (see e.g. Tielens 2005); although this must be balanced against the tendency of such shocks to sputter and vaporize the grains (Jones et al. 1996).

All of these mechanisms, together with thermal emission from shocked ionised gas, and synchrotron emission from within the SNR shells themselves, suggest that the regions may be capable of being detected over broad ranges of IR wavelengths. It is also clear that interaction with the ISM is likely to transform the morphology of local interstellar material, and lead to fragmented and wind-swept features such as have been discussed by Phillips et al. (2009).

We shall present MIR observations of a further four SNR, based upon imaging undertaken during the Galactic Legacy Infrared Midplane Surveys Extraordinaire I, II and 3D (GLIMPSE I, II, & 3D; referred hereafter as GLIMPSE). This work therefore extends upon the previous GLIMPSE I survey of these sources undertaken by Reach et al. (2006). We shall find, as for the latter investigation, that there is little evidence for SNR shells to the limiting sensitivity of the Spitzer Space Telescope (hereafter referred to as *Spitzer*). However, several examples are discerned of interaction between SNRs and the ISM, and we consider these cases in the following discussion. We have also undertaken an analysis of possible triggered Class I YSOs within W28, fitting 2D radiative transfer modelling of young stellar objects (YSOs) to the observed NIR-MIR spectral energy distributions (SEDs). We shall find that the YSOs are, as expected, likely to have broad ranges of luminosity and mass, although tighter constraints should be possible where the surveys are deeper, and samples are larger.

## 2. Observations

We shall present photometry and imaging of four Galactic SNRs deriving from the near infrared (NIR) 2MASS all sky survey, and the MIR Galactic Legacy Infrared Midplane Survey Extraordinaire (GLIMPSE); the latter of which programmes was undertaken using *Spitzer*. Details of the data bases employed, and procedures used in the analysis of the data can be found in Skrutskie et al. (2006) (2MASS), and Benjamin et al. (2003) (GLIMPSE). We shall describe here only the most salient features of importance to this analysis.

The GLIMPSE survey mapping was undertaken in three primary phases denoted as GLIMPSE I, II AND 3D, and used the Infrared Array Camera (IRAC; Fazio et al. 2004), and filters having isophotal wavelengths (and bandwidths $\Delta\lambda$) of 3.550 $\mu$m ($\Delta\lambda$ = 0.75 $\mu$m), 4.493 $\mu$m ($\Delta\lambda$ = 1.9015 $\mu$m), 5.731 $\mu$m ($\Delta\lambda$ = 1.425 $\mu$m) and 7.872 $\mu$m ($\Delta\lambda$ = 2.905 $\mu$m). The spatial resolution varied between ~1.7 and ~2 arcsec (Fazio et al. 2004), and is reasonably similar in all of the bands, although there is a stronger diffraction halo at 8 $\mu$m than in the other IRAC bands. This leads to differences between the point source functions (PSFs) at ~0.1 peak flux. The maps were published at a resolution of 0.6 arcsec/pixel.

We have used the Spitzer GLIMPSE results to produce imaging at 5.8 and 8.0 $\mu$m; fluxes were weaker in the 3.6 and 4.5 $\mu$m bands, and the sources barely detected these channels. We have therefore not included imaging at the shorter IRAC wavelengths. We have also used the GLIMPSE point source catalog (PSC), together with the diagnostic diagram of Allen et al. (2004) to identify likely Class I YSOs close to the SNRs.

Surface brightness profiles have been obtained in all four of the IRAC bands, and are composed of "background" components of flux (mainly dominated by PAHs), components arising from more localised cloud structures, and emission deriving from the fronts associated with SNR/ISM interaction. We have removed a constant value of surface brightness from each of the profiles in order to emphasise changes associated with the fronts themselves. The observed surface brightness S(3.6) at 3.6 $\mu$m is represented in terms of S(3.6) = $S_{REV}$(3.6) + $\Delta$3.6, for instance, where $S_{REV}$(3.6) is the relative surface brightness illustrated in our profiles, and $\Delta$3.6 is the background component of emission. The values ($\Delta$3.6, $\Delta$4.5, $\Delta$5.8, $\Delta$8.0) are indicated in the captions to the figures. It is pertinent to note that since surface brightnesses change so markedly over the fields, it has been necessary to estimate $\Delta$3.6, $\Delta$4.5 & etc. at locations close to the frontal structures; a process which involved taking estimates of "background" at five positions about the regions, and within 2 arcmins of the areas under investigation. This does not, of course, give a guaranteed measure of frontal flux, excluding all other contributions and possible contaminating sources, and it is, indeed, likely to lead to underestimates in intrinsic brightnesses. Having said this, however, this represents one of the more reliable procedures that one could adopt, and is considerably more trustworthy than using uncorrected values of S.

We have finally used values of $S_{REV}$ to determine the flux ratios $F(3.6\mu m)/F(5.8\mu m)$ and $F(4.5\mu m)/F(8.0\mu m)$, although this is only possible where the 3.6 and 4.5 $\mu m$ fluxes permit viable estimates of these parameters. These have been used to assess likely emission mechanisms within the clouds using the diagnostic colour diagram of Reach et al. (2006). Such a flux ratio analysis is necessarily somewhat approximate, given that estimates of $\Delta 3.6$, $\Delta 4.5$ & etc may be slightly in error; that shorter wave fluxes may be contaminated by stars; and that surface brightnesses S are affected by scattering in the IRAC camera. This latter problem has been described in several previous papers (see e.g. Phillips et al. 2010), and may require corrections of as large as 0.944 at 3.6 $\mu m$, 0.937 at 4.5 $\mu m$, 0.772 at 5.8 $\mu m$ and 0.737 at 8.0 $\mu m$ (see the IRAC Data Handbook at http://ssc.spitzer.caltech.edu/irac/dh/iracdatahandbook_3.0.pdf). All of these factors, together with the uncertainties associated with extinction, may affect the positioning of sources within the $3.6\mu m/5.8\mu m$-$4.5\mu m/8.0\mu m$ colour plane – and lead to errors in evaluating the nature of the emission arising at the fronts. The possible consequences of such uncertainties are described in Sect. 3.

The 2MASS all-sky survey was undertaken between 1997 and 2001 using 1.3 m telescopes based at Mt Hopkins, Arizona, and at the CTIO in Chile, and was used to provide all sky photometry and imaging in the J (1.25 $\mu m$), H (1.65 $\mu m$) and $K_S$ (2.17 $\mu m$) photometric bands. We have used photometry from the associated PSC to evaluate SEDs for 11 YSOs, and employed these in modelling star formation in W28, as described in Sect. 4.

## 3. MIR Imaging of SNR showing evidence of Interaction with the ISM

### 3.1  SNR G001.0-00.1 (Sgr D)

The region Sgr D is composed of two emission components; a northerly HII region, and the more southerly SNR G001.0-00.1 (see e.g.

the 90 cm VLA image of LaRosa et al. (2000), the 18 cm results of Liszt (1992), and the 6 cm mapping of Mehringer et al. (1998)). These are illustrated in Fig. 1, where the radio mapping from LaRosa et al. (2000) is indicated as green, and emission to the upper left of the panel corresponds to the HII region. The SNR has non-thermal emission at radio wavelengths (spectral index $\alpha$ = -0.4; LaRosa et al. 2000), has been detected at X-rays by Sidoli et al. (2001) and Ryu et al. (2009), and appears to be associated with a compact OH 1720 MHz maser spot (Yusef-Zadeh et al. 1999), normally taken to be an indication of interaction between SNRs and the ISM (see e.g. Yusef-Zadeh 2003). Finally, we note that the distance and Galactic coordinates of the source place it close to or just beyond the Galactic centre (Downes et al. 1980; Odenwald & Fazio 1984; Liszt 1992; Gray 1994b; Mehringer et al. 1998).

A superposition of the radio results cited above, the thermal 1720 MHz mapping of Yusef-Zadeh (1999) (white contours); several Class I YSOs (blue circles) and our present 8.0 $\mu$m Spitzer imaging (red) is illustrated in Fig. 1. The YSOs are identified using [3.6]-[4.5] and [5.8]-[8.0] colours deriving from the PSC, allied with a theoretical diagnostic diagram due to Allen et al. (2004). Several interesting characteristics are immediately apparent.

It is clear for instance that the SNR shell appears to occupy a cavity-like region within the MIR emission regime. This suggests that the SNR and ISM are likely to be interacting, leading to the removal of pre-existing interstellar material and/or the destruction of PAH emitting grains. There is, apart from this, also evidence for YSO sources close to the lower limits of the SNR shell (where point source maser emission has also been detected, as noted above); at other positions close to the periphery; and in association with the central (lower radio intensity) lane (viz. the three point sources close to $l$ = 1.04°, $b$ = -0.17°). This may represent the clearest evidence for triggered star formation in any of our sources.

Finally, we note evidence for shock interaction with small clouds or inhomogeneities in the local ISM. A condensation to the lower left-hand side of Fig. 1, for instance, shows evidence for possible shock interaction with the SNR shell. Not only is it located at the precise periphery of the shell, but it has the convex frontal structure and swept-back arms to be expected where the SNR is shocking the source, and ram-pressure stripping material from the globule.

The structure of this feature is more clearly seen in Fig. 2, where we have represented the cloud at various levels of imaging saturation. The position of the globule vis-à-vis the main SNR shell is for instance most clearly to be seen in the upper panel, where it is apparent that the SNR shell engulfs frontal portions of this structure. The insert to this figure, by contrast, has a significantly higher level of saturation, and shows the cometary shape defined by outer (and fainter) portions of the structure.

Finally, the lower panel shows MIR profiles through the centre of the globule, and both this and the inserted image reveal how emission is concentrated in a rim. Such a structure may arise where the external wind impacting the cloud is leading to shocked neutral and/or ionized gas.

Finally, I($3.6\mu m$)/I($5.8\mu m$) and I($4.5\mu m$)/I($8.0\mu m$) flux ratios for the profile are illustrated in Fig. 3, where we also define various emission regimes based upon the analysis of Reach et al. (2006). We also indicate the reddening vector for an extinction $A_V$ = 20 mag, and adopting the IS reddening curve of Indebetouw et al. (2005), from which it is clear that any appreciable level of extinction would cause a vertical shift downwards in the ratios.

Although extinction in this, and in most of the other SNRs is very poorly known, the ratios for this present SNR are probably least affected by stellar contamination, and likely to give a reasonable indication of the primary emission mechanisms. Our results therefore suggest that fluxes are being enhanced through the shock excited v = 0→0 transitions of $H_2$. Note that the CO fundamental band is also strong in Herbig-Haro shocks, and would be expected to increase 4.5 $\mu m$ fluxes, and shift sources to the right (Reach et al. 2006). This mechanism cannot entirely be discounted for the regions investigated here.

It is therefore clear that G001.0-00.1 is an extreme example of a SNR which, despite its roughly circular morphology, appears to be directly interacting with the ISM.

## 3.2 G 355.9-02.5

The SNR G355.9-02.5 was first detected through the 408 MHz and 5 GHz observations of Clark et al. (1975), and later mapped at 843 MHz using the Molonglo Observatory Synthesis Telescope (MOST) (Caswell et al. 1983; Gray 1994b), and at 1465 MHz by Dubner et al. (1993). The source has an angular diameter of 13' ± 0.8' (Dubner et al. 1993), and a spectral index $\alpha \cong -0.5$ which appears to be more-or-less constant over the source (Clark et al. 1975; Dubner et al. 1993; Kassim 1989). Dubner et al. (1993) find 6 % linearly polarized emission (particularly strong in the bright southerly and westerly portions of the shell), whilst there is little evidence for any associated infrared (Arendt 1989; Saken, Fesen & Shull 1992) or 1720 MHz maser emission (Green et al. 1997). The $\Sigma$-D calibration of Guseinov et al. (2004) places the source at a distance of 7.4 kpc, the calibration of Case & Bhattacharya (1998) gives 7.6 kpc, whilst Dubner et al. (1993) estimate $D \cong 8.0$ kpc.

Caswell et al. (1983) noted the strongly asymmetric structure of the shell, and the fact that whilst the SE portions of the envelope (i.e. those located towards higher Galactic latitudes l) are more intense, those to the west (lower l) are very much weaker. The latter trend may arise where the ISM is relatively hotter and of lower density, leading to a so-called "blow-out" of the shell. Gray (1994b) also noted that the SNR displays unusual linear features to the east – structures which appeared to be detached from the main SNR shell, and perhaps "related to structure in the surrounding interstellar medium".

These latter features can be noted to the lower left of Fig. 4, where we have superimposed Gray's 843 MHz MOST map (in green) upon the GLIMPSE 8.0 $\mu$m imaging (in red). The position of the Class I YSOs is indicated by yellow circles.

It is clear that the lineal features of Gray are at the edge of a parallel region of 8.0 $\mu$m emission; the radio features appear to follow, and fall just outside of the 8.0 $\mu$m structural components. It is therefore likely that expansion is being modified by the ISM, as previously surmised by Gray. It follows that the partial fragmentation of the easterly shell is unlikely to arise through the putative blow-out described by Caswell et al. (1983), but as a result of direct interaction with the ISM. A problem, in this latter case, would be to explain why levels of radio surface brightness are so relatively modest; the reverse of what occurs in other regions of SNR/ISM interaction (see e.g. Sect. 3.4). Although we have no ready-made answer to this interesting (and troubling) question, it is possible that the regime of interaction in the SNR is highly evolved. Where this is the case, then levels of post-shock cooling may be considerably reduced, leading to low fractional ionisation of the SNR shell.

Even more direct evidence for interaction is to be noted in the lower portions of this figure, where we see several parabolic arcs associated with likely frontal structures. It is probable, in this latter region, that the SNR is interacting with compact ISM clouds or inhomogeneities, leading to triggered star-formation at the limits of these structures. Note the three YSOs positioned close to the edges of the two principal arcs. These arcs are also illustrated in the lower left-hand panel of Fig. 4, where we see evidence for a narrow structure with apex close to $l = 355°\ 54'\ 11$, $b = -2°\ 37'\ 39$ (this is the weakest of the features); a broad feature (size ~7.6 arcmin) with apex close to $l = 355°\ 55'\ 13''$, $b = -2°\ 38'\ 30''$; and the bright region of interaction at $l = 355°\ 59'\ 15$, $b = -2°\ 38'\ 57''$, close to which there is also evidence for a Class I YSO. It should be emphasized that although we are clearly seeing evidence for ISM-SNR interaction, the radio surface brightnesses in this region are again relatively weak.

Profiles through two of the frontal structures – that associated with the bright narrow ridge towards lower Galactic longitudes (profile A), and the other close to the apex of the broadest structure (profile B) are illustrated in Fig. 5. These show certain characteristics in common – but also possess several important differences.

It is apparent for instance that the frontal structure A is very well defined, has a FWHM at 8.0 $\mu$m of ~10 arcsec, and appears to have an asymmetric profile, such that emission rises steeply at the leading edge of the shock (in the region -5 arcsec < RP < +5 arcsec), and tails off more slowly to negative RPs. The latter, slower fall-off may presumably arise as a result of cooling within a post-shock cooling regime; through curvature of the shock along the line of sight; and/or through Kelvin-Helmholtz and Rayleigh-Taylor instabilities, leading to complex, extended, and asymmetrical frontal structures. The front is also evident at 5.8 µm, but very much weaker at shorter wavelengths - if, indeed, it is present at all. All of these trends will be seen to be very similar to those observed for other sources in this study.

A putative Class I YSO occurs close to the centre of the ridge, at the position of peak 8.0 µm emission, and is clearly evident as a point source component in the 3.6 and 4.5 µm results. It is far from clear, however, whether there is any corresponding 5.8 and 8.0 µm emission (as supposed in the GLIMPSE PSC), or whether the longer wave fluxes derive from the frontal structure alone. Although this source therefore represents one of the more promising examples of triggered star formation, it remains uncertain whether its classification is entirely to be trusted.

The trend for increased longer wave emission is also to be noted for frontal structure B (lower panel of Fig. 5), although in this case, the emission rises strongly at the front, and then stays more or less constant to negative RPs. There are various ways in which such a trend may arise, including where the longer wave emission arises from a pre-existing IS cloud, with little contribution from shock compression and heating. It is also possible (although less likely) that we are seeing a very extended post-shock cooling regime; emission from foreground and background shock emission fronts; and the influence of multiple fronts (including reverse shocking of the gas).

We finally note the presence of two strong, and unresolved regions of emission at shorter IRAC wavelengths – regions which are located close to RP = -2.8 and +1.6 arcsec, but appear not to be present at 8.0 µm, or are merged within the broader emission peak. These may represent Class II-III YSOs close to the HI/HII interface.

Finally, it will be recognized that the 3.6 and 4.5 µm fluxes are heavily compromised by a skein of field stars within the image - sources which are mostly very faint at these wavelengths, but capable of strongly affecting the shorter wave fluxes. It is very difficult, under these circumstances, to come to hard-and-fast conclusions concerning shell emission mechanisms. It is also important, in such analyses, that one removes the strong background emission arising from unrelated

Galactic emission, as we have done for the profiles illustrated in this paper; failure to do so fatally affects deduced flux ratios.

Despite these caveats, the flux ratios for slice A are indicated in Fig. 3 (denoted as G355.9-02.5 A), based on mean fluxes for the central portions of the profile (-11 < RP < 4 arcsec). The result is consistent with $H_2$ excitation and possible 4.5 $\mu$m CO emission, as was previously noted for G001.0-00.1 (Sect. 3.1). Given the likely level of stellar contamination, however, it is entirely possible that we are witnessing enhanced PAH emission, or a combination of all three emission processes ($H_2$ + CO + PAH).

### 3.3 G355.6-00.0

The SNR G355.6-00.0 has a roughly spheroidal morphology in the 843 MHz MOST maps of Gray (1994a), although there is also some evidence for a protuberance to higher Galactic latitudes which may arise due to a blow-out of the shell (see e.g. Fig. 6, where the radio emission is represented as green). By contrast, there is also evidence, in the original maps, for enhanced surface brightnesses towards the upper portions of the structure (higher Galactic latitudes), where the shell is seen to abut a region of diffuse ionised emission. Although Gray noted that the source appears not to have been detected in either the far infrared (IRAS), or in the Parkes radio data set, suggesting that much of the flux may derive from synchrotron emission, he also notes that diffuse emission in the upper portions of Fig. 6 is likely to arise from thermal components of flux. Finally, the distance to the SNR is very poorly determined, although Guseinov et al. (2004) suggest a value of 12 kpc, and Case & Bhattacharya (1998) find 12.6 kpc, both of which estimates are based upon $\Sigma$-D calibrations.

The superposition of the radio data of Gray (1994a,b) upon our GLIMPSE 5.8 and 8.0 $\mu$m imaging is illustrated in Fig. 6, where we have represented 5.8 $\mu$m emission as green, and 8.0 $\mu$m fluxes as red. It is apparent, from flux ratios, that most of the diffuse emission within the field derives from the 6.2, 7.7 and 8.6 $\mu$m PAH emission bands, although individual compact HII regions (particularly that located close to $l$ = 355° 33' 30", $b$ = -00° 05' 51") have stronger shorter wave emission, and a yellow-green appearance. The Class I YSOs are indicated by the blue circles, and it is apparent that they cover the entire area of the image, with no particular concentration near HII regions or HI/HII interfaces. It therefore follows that whilst some of the sources may be close to or co-spatial with the SNR, there is little evidence for triggering of star formation by G355.6.0-00.0.

There are however certain indications for possible interaction between G355.6-00.0 and the surrounding gas, and these take the form of enhanced levels of extinction, and a possible shock interface towards the west (i.e. towards the upper centre of Fig. 6).

To take the case of the enhanced extinction first, it is noteworthy that the lower boundary of the SNR shell appears to be associated with a narrow band of obscuration – a band which may presumably arise as a result of the snow-plough accumulation of material at the leading edge of the SNR shell. Further, rather more obvious, but also more ambiguous evidence for such an interaction may also be noted towards the east of the source, in the upper regions of the image in Fig. 6. It would seem that an arc of extinction is surrounding the upper portions of the diffuse emission shell, with some evidence for a diminution (or gap) in extinction close to $l = 355°$ 43' 20", $b = +00°$ 04', where a tongue of ionized emission also extends upwards. That some kind of interaction with the ISM and the diffuse gas is occurring in this zone, and is perhaps responsible for compressing material ahead of the HI/HII interface, is also suggested by the rim of yellow-coloured emission close to $l = 355°$ 41' 42", $b = 00°°$ 02' 50"; a feature which is more clearly seen in the expanded version of this region illustrated in the lower panel of Fig. 6.

It is unclear, in this case, whether we are observing interaction of precursor supernovae winds with the surrounding ISM, and that these are responsible for both the diffuse westerly emission and possible shocking with the ISM; whether this latter emission is simply an unrelated line of sight HII region; or whether the diffuse emission represents a larger and fainter portion of the main SNR shell, perhaps deriving from a blow-out towards higher Galactic latitudes.

The geometry of the structure, however, and the way in which the partial arc of extinction, and associated yellow-enhanced emission appear to be centred upon the SNR, suggests that these features may be a direct consequence of interaction between the ISM and G355.6-00.0.

Profiles through the probable frontal structure are illustrated in Fig. 7, where the directions of the slices are indicated in the lower panel of Fig. 6, and in the inserted panels in Fig. 7. Both of these show almost identical trends at 3.6 and 4.5 $\mu$m, and indicate that fluxes in these bands are strongly affected by field star contributions. The profiles also reveal a strong increase in surface brightnesses towards 5.8 and 8.0 $\mu$m, as noted for other sources in this study (viz G001.0-00.1, G355.0-02.5). Finally, the positions of the fronts are clearly defined by the humps of emission close to RP = 0 arcsec, whilst complexity in these structures, possibly caused by instability and curvature of the fronts, causes broadening of the emission region by ~1 arcmin. The primary 5.8 and 8.0 $\mu$m emission, however, is clearly concentrated within a range $\Delta$RP ~ 20-40 arcsec.

The profile in the upper panel (slice A) also crosses the position of one of the supposed YSOs (at RP = 0 arcsec), for which it would seem that peak emission occurs in the 5.8 $\mu$m channel.

We have finally indicated the flux ratios for these profiles in Fig. 3, where slices A and B are indicated respectively as 355.6-00.0A and 355.6-00.0B. We have again attempted to estimate fluxes for regions which are least contaminated by field star emission – although this does not guarantee that stellar components are entirely negligible. On the contrary, it is possible that the 3.6 and 4.5 $\mu$m bands are indeed affected by such continua, and seriously modify the ratios presented in Fig. 3.

The ratios for slice B are consistent with enhanced components of (shock enhanced?) $H_2$ emission (see Fig. 3) and the 4.5 $\mu$m CO fundamental band, whilst it is conceivable that profile A, which lies between the molecular and ionised regions of Fig. 3, is also affected by [FeII] $\lambda$ 5.34 $\mu$m emission within the 5.8 $\mu$m band – a transition which tends to drag 3.6$\mu$m/5.8$\mu$m ratios to lower positions within the plane. It is also however clear, as noted above, that stellar emission may have a role in increasing these ratios by factors of > 2. Thus for instance, we note that relative fluxes for PAH band emission are of order 0.04/0.046/0.35/1.0 for IRAC bands 1/2/3/4 (Reach et al. 2006), and that stellar fluxes fall-off as 4.9/3.1/1.9/1.0. it follows that where the stellar contribution at 3.6 μm accounts for half of the measured flux, and PAH bands for the other half, then the ratios F(3.6μm)/F(5.8μm) and F(4.5μm)/F(8.0μm) would increase to ~0.22 and ~0.7. This would place the ratios at the limits of the molecular emission regime illustrated in Fig. 3. It is therefore clear that where stellar contamination is appreciable, then the role of PAH emission cannot be discounted.

### 3.4 G006.4-0.1 (W28)

The SNR W28 represents an example of the so-called "mixed morphology" or "X-ray composite" sources first defined by Rho & Petre (1998) and Jones et al. (1998), whereby the radio morphology is consistent with a shell-like structure, whilst X-ray emission is concentrated towards central portions of the source. Various observations have confirmed that the X-ray emission in W28 is thermal in nature, a tendency which is also shared by other mixed morphology sources (see e.g. the Einstein IPC observations of Long et al. (1991), the analysis of ASCA data by Torii et al. (1993), and analyses by Rho et al. 1994; Rho 1995; Rho & Petre 1998). It has been suggested that such X-ray structures arise as a result of fossil radiation from the shell formation (radiative stage) of evolution (Cox et al. 1999; Chevalier 1999), or through the evaporation of interior clouds (White & Long 1991). Neither of these models appears entirely satisfactory when it comes to explaining the emission in W28, however (Rho & Borkowski 2002). Radio mapping shows that the shell structure is irregular, with the highest levels of surface brightnesses occurring to the N and NE (e.g. Kundu 1970; Shaver & Goss 1970; Milne & Wilson 1971; Kundu & Velusamy 1972; Dubner et al. 2000), where it has been suggested that the SNR is interacting with the ISM. Evidence for such interaction also comes from the more than 40 discrete 1720 MHz maser spots detected in this region (Hewitt, Yusef-Zadeh & Wardle 2008; Hoffman et al. 2005; Claussen et al. 1997); the presence of shocked $H_2$ and ionic lines detected using the Infrared Space Observatory (ISO) (Reach & Rho 2000), *Spitzer* (Neufeld et al. 2007), and through visual observations (Long et al. 1991; Bohigas 1983), results which imply shock velocities in excess of 35 km s$^{-1}$; and from the presence of high velocity CO J = 1-0 and 3-2 emission deriving from post-shock gas (Arikawa et al. 1999; Frail & Mitchell 1998; Wootten 1981). The source also appears to be associated with an enveloping HI shell, presumably corresponding to IS material swept up by the SNR shell (Velazquez et al. 2002).

Extinction towards the source has been measured as being E(B-V) = 1.16 (Bohigas 1983), and E(B-V) = 1 – 1.3 (Long et al. 1991), whilst estimates of distance based upon radio $\Sigma$-D relations imply a range 1.6 to 3.0 kpc (Clark & Caswell 1976; Milne 1979; Frail et al. 1994); values based upon $\Sigma$-D relations and optical measurements suggest that D $\cong$ 1.8 – 2.0 kpc (Goudis 1976 and Long et al. 1991); and observations of molecular line broadening imply a value 1.9 ± 0.3 kpc (Velazquez et al. 2002). We shall adopt the values of D of Velazquez et al. in our later analysis of YSO sources (see Sect. 4).

Certain of these emission components are apparent in Fig. 8, where we have superimposed the 0.5-2.4 keV Rosat X-ray image of Rho & Borkowski (2002) (blue), and the 1415 MHz VLA radio mapping of Dubner et al. (2000) (green) upon the MIR image deriving from the GLIMPSE survey; where in the latter case 5.8 $\mu$m emission is represented as blue, and 8.0 $\mu$m emission is red. The white contour represents the approximate limits of the SNR shell based on the 1415 MHz observations of Dubner et al. (2000), for which the limiting background noise levels are of order ~4.3 mJy arcmin$^{-2}$. Finally, we have again used *Spitzer* photometry of point sources in the region, allied with the colour-colour modelling of Allen et al. (2004) to define the positions of some 190 Class I YSOs within the field of W28 (indicated by blue circles). Although a proportion of these sources may be misclassified, and correspond to Class II YSOs (see Sect. 4), it is clear that all of them represent early stages of stellar evolution.

Several aspects of the figure are worth remarking upon in the context of the results summarised above. The first is that the morphology of this source is by no means straightforward, and we are seeing the two-dimensional projection of a complex three dimensional outflow. There is therefore plenty of opportunity for misinterpreting the structure. It is nevertheless apparent that the NE portion of the source has much higher levels of surface brightness, and that this defines the primary region of interaction with the ISM. The low surface-brightness, south-westerly (blow-out?) region may also show evidence for interaction with the ISM. We note for instance a parabolic shock-like feature close to $\alpha(2000.0)$ = 17:59:52, $\delta(2000.0)$ = -23:45:16, probably indicating an interaction between the SNR shell and a pre-existing IS globule.

There is also evidence for a narrow and filamentary frontal structure snaking down the extreme left-hand side of the figure, and curling horizontally along the lower parts of the image; a feature which is clearly well outside of the main SNR shell. The way in which this structure appears to partially wrap around the SNR is extremely suggestive, and may imply an earlier phase of interaction between the ISM and pre-cursor SN winds.

The Class I YSOs extend throughout the field, although their spatial densities within the SNR appear significantly greater than for adjacent regions of sky. The YSOs within the SNR, for instance, have a spatial density which is ~3.4 times greater than for regions immediately to the left (i.e. to larger right ascensions in Fig. 8; ~124 deg$^{-2}$ over an area of 0.45 deg$^2$, compared to ~423 deg$^{-2}$ over the same area of W28). A more general survey of regions on all sides of the SNR, but excluding the Triffid nebula to the north, suggests that the ratio is closer to ~ 2.1 (i.e. the region outside of W28 has 204 Class I YSOs per square degree, over an area of sky which is double that of W28). It follows that between a half and two thirds of stars within the borders of W28 may have been triggered through interaction between the SNR and local ISM.

The northerly interaction region of the W28 shell is further illustrated in the upper panel of Fig. 9, where apart from presenting a two-colour MIR image, we also show the distribution of 327 MHz emission (white contours; Frail et al. 1993) and indicate regions of higher velocity CO J = 3-2 gas (green borders; Arikawa et al. 1999). The white circles mark the locations of the Class I YSOs. Two things are apparent from this image. The first is that the upper area of shocked CO also coincides with a ridge of MIR emission. The dark bar across this feature indicates the directions and widths of the slices illustrated in Fig. 10 (upper panel). It is clear, from the latter profiles, that the ridge is strong at 5.8 and 8.0 $\mu$m, almost invisible at 3.6 and 4.5 $\mu$m, and has a roughly symmetric profile centred close to RP = 0 arcsec. We have also acquired similar profiles across the bright ridge associated with the southerly region of high velocity CO gas (the lower panel in Fig. 10). The flux ratios for these profiles (W28A and W28B) appear to imply that much of the emission may derive from shocked $H_2$ transitions (see Fig. 3).

By contrast, the unrelated southerly region of interaction is illustrated in the lower panel of Fig. 9, and appears to represent a zone of flocculent 1415 MHz radio emission. The profile through the ridge is illustrated in Fig. 11, and shows the typically sharp-peaked asymmetrical structure noted in certain of the other fronts investigated within this study (see e.g. G355.9-02.5; G355.8-00.0), whereby one side of the profile rises steeply, and the other side declines more gradually. Given the morphology of the arc, and assuming that the SNR is impacting from the north, then it may be possible that the right-hand limits of the profile (at RP ~ 90 arcsec) represent the position of a shock front which is entering a dark, highly extincting globule, whilst the more extended emission (between RP < -50 and 50 arcsec) corresponds to less obscured regions of background emission.

It seems clear, from Fig. 3, that the flux ratios imply enhanced PAH emission (the relevant symbol is referred to as W28C). This may suggest that any associated shock has been sufficient to shatter pre-existing grains, and generate and excite a larger number of small PAH emitting particles, or that there are other local sources of radiation which are capable of exciting the PAH bands.

## 4. Discussion

It is clear, from the foregoing analysis, that there is multiple evidence for interaction between the ISM on the one hand, and the present SNRs on the other. These include the presence of likely shock fronts where outflows collide with interstellar gas, leading to parabolic structures (W28, G001.0-00.1, G355.9-02.5, G355.6-00.0), and apparent voids in the MIR emission (G355.9-02.5, G001.0-00.1). They also include possible triggering of star formation in G001.0-00.1, leading to a necklace of YSOs close to the perimeter of the shell; in G355.9-02.5, where YSOs are located along frontal arcs; and in W28, where the spatial density of YSOs appears to have been locally enhanced.

Where the profiles of the fronts have been investigated in this study, then it seems that there is a strong increase in 5.8 and 8.0 $\mu m$

emission, but very little evidence for corresponding shorter wave enhancement. Similarly, we note that several fronts show evidence for strong increases in emission to one side of the structures, followed by a slower decline to the other – a morphology to be expected where there is gradual cooling within a post-shock regime and/or curvature of the front along the line of sight.

Almost all of the profile flux ratios I(3.6$\mu$m)//5.8$\mu$m) and I(4.5$\mu$m)/I (8.0 $\mu$m) appear consistent with shock excitation of the v = 0$\rightarrow$0 transitions of $H_2$. This trend may, however, turn out to be partially illusional. We note for instance that emission at 3.6 and 4.5 $\mu$m is often contaminated by stellar continua, and this, together with PAH band emission, can lead to ratios which are similar to those observed here. This is less the case in the sources G001.0-00.1 and W28, however, where levels of field star and YSO contamination appear less. Similarly, Reach et al. (2005) note that the Fe II $\lambda$1.644 $\mu$m transition is present in W28, although levels of emission appear weak, and the spatial distribution is relatively diffuse. Given that flux ratios for G355.6-00.0 are also consistent with emission by [FeII], then it would appear that this transition may also contribute to the observed MIR fluxes. It would therefore seem that the least one can say concerning the present MIR fluxes is that $H_2$ emission, PAH band emission, and ionic transitions such as [FeII] are all likely to play a role, and are capable (to varying degrees) of explaining the present results.

Where triggered star formation occurs, then it is possible to use YSO modelling of the SEDs to constrain the physical properties of the stars. We shall undertake such an analysis for 103 Class I YSOs located within the region of interaction of the W28 shell (indicated using white circles in the upper panel of Fig. 9). Of these, some 83 % (a total of 86) had only GLIMPSE IRAC photometry, and this was insufficient to adequately constrain the model parameters. A further 6 sources had a full range of IRAC and 2MASS photometry, with the SED of one of the sources also being defined through MSX photometry; however, the ages of the models were in excess of the lifetime of the SNR shell. This leaves only 11 YSOs having adequate photometry and ages consistent with formation by the supernova event – a very small fraction of the total which were probably triggered by the SNR. These sources are indicated using blue crosses, and are located close to the CO post-shock regime.

It is worth emphasising that the YSOs in W28 were selected using the modelling results of Allen et al. (2004), in which the differing classes of YSO are identified by means of their MIR colours. However, this classification of YSOs (in terms of Class I, II & III sources) represents an imperfect way of determining their state of evolution; differing orientations of protostars can lead to differing estimates of Class, even though their phases of evolution are very closely similar. It has therefore been proposed that YSOs are better classified in terms of an evolutionary "stage", defined to accord more closely with their actual states of evolution.

An example of such an analysis is provided by Robitaille et al. (2006), whence it is apparent that the use of [3.6]-[5.6] and [8.0]-[24.0] indices permits a clearer discrimination between the "stages" of evolution. Stage I YSOs (roughly the equivalent of Class I sources) have significantly larger [8.0]-[24.0] indices than is the case for Stage II, for instance, and such an analysis would be more appropriate (and less ambiguous) than procedures based upon the modelling of Allen et al. (2004). However, point source photometry is still not available for the MIPSGAL survey, and the number of 24 $\mu$m sources within W28 is of order $\sim 2.5 \times 10^3$; a total which makes the estimation of these fluxes extremely time-consuming. We have therefore limited ourselves to estimating Class I sources using the diagnostic diagram of Allen et al. (2004).

The location of Class I sources in the primary shell of W28 (Fig. 8), and of the candidate YSOs in the region of interaction (Fig. 9), are identified through the respective blue and red symbols in Fig. 12. We also show the reddening vector for $A_V = 20$ mag (based on the extinction trends of Indebetouw et al. (2005)), and mean errors in the YSO colour estimates; although it should be noted that actual ISM reddening is $\sim$5-6 times smaller (see Sect. 3.4). It will finally be noted that we have identified sources having indices [5.8]-[8.0] > 1, [3.6]-[4.5] < 0.3 as Class I YSOs, although they extend somewhat below the modelling of Allen et al. (2004); it is likely that they represent Class I sources having lower envelope densities. Similarly, a small proportion of sources have larger indices [5.8]-[8.0] and [3.6]-[4.5], and lie above the regime of the Class I models. These are also likely to correspond to Class I sources, and are so interpreted here. Their location outside of the Class I regime may be due to errors in the photometry, or a consequence of the limited ranges of YSO parameter used by Allen et al. The inclusion of these sources makes little difference to our conclusions or results.

It is clear that given the uncertainties in the colours, then a fraction of the sources may represent Class II YSOs – although it is likely the majority of the sample to be analysed below (red circles) are indeed Class I stars.

It is finally worth adding that W28 is the only source for which the distance D, extinction $A_V$, and age $T_{EV}$ are reasonably well constrained (see the references cited in Sect. 3.4 as well as those mentioned below) – factors which permit a tolerably realistic analysis of the associated YSOs.

We have evaluated infrared SEDs for all of these YSOs using photometry deriving from the GLIMPSE and 2MASS surveys. These have subsequently been analysed using the 2D radiative transfer modelling of Robitaille et al. (2006), the fitting tool of Robitaille et al. (2007), and assuming that $3 < A_V/\text{mag} < 4$ and $1.6 < D/\text{kpc} < 2.2$ (see Sect. 3.4).

Some of results from our modelling are illustrated in Fig. 13, and apply for a YSO located at l = 6.535°, b = -0.1009°. The graphs show the variation in several physical parameters against the lifetime of the star, where filled bullets correspond to the best fit models (those having $\chi^2 - \chi_{BEST}^2 < 3$), and grey areas represent the ranges of parameter explored. If one assumes that the formation of the stars was triggered by the W28 SNR, then we can establish that the age of the YSO cannot be greater than $3.3 - 3.6 \times 10^4$ yrs (Rho & Borkowski 2002; Velazquez et al. 2002). This implies that masses lie in the range $\Delta M \sim 0.35\ M_\odot$, rather than the range $\Delta M \sim 4.75 M_\odot$ implied by all of the model fits taken together. Note that the age of the SNR is likely to represent an upper limit for the lifetime of this and other sources in this study. However, where stellar ages are different by factors of up to two or three, then this makes little difference to the following results.

More generally, where we now define a range of model masses $\Delta M_i$ for any particular YSO *i*, one can then represent it in a histogram with height $\Delta M_i^{-1}$, centred upon the mean model mass $M_i$. Where we follow this procedure for all of the YSOs, then one can build up a mass spectrum in which each YSO is given an equal weighting.

This procedure has been followed for a broad range of parameters, and the results are illustrated in Fig. 14. All of the histograms have been normalized to unity.

It is clear from this that the physical parameters of the stars extend over broad ranges of value, and that there is no particular bias towards higher or lower values. This may, at first sight, seem to be somewhat surprising, since the initial mass and luminosity functions of such stars are likely to be biased towards lower values of the parameters (i.e. there should be many more lower mass and luminosity YSOs). Why should there be this discrepancy between theoretical expectations and our present results?

One problem with this analysis is that where the stars have not have been triggered by W28, then our constraints upon the stellar ages may turn out to be spurious. However, it seems likely that the higher spatial densities of sources within W28 implies that a high proportion of the YSOs were formed through triggering by the SNR (see Sect. 3.4). When this is combined with the location of the candidate sources within the region of interaction, and the apparent ability to represent the SEDs of these stars using low $T_{EV}$ YSO models, then it seems likely that a good fraction of the candidate stars will have been triggered by the SN event.

A further problem is that although the sample size in W28 is larger, it is still relatively modest. This may result in the well known biases associated with limited sample sizes, whereby a few unrepresentative sources are capable of causing untypical parametric trends.

A further and very important caveat relates to the probability of seeing differing masses of star in the MIR. Whilst high mass YSOs evolve extremely rapidly, and represent a small fraction of the stellar content, they are also very much brighter and more easy to detect. By contrast, lower mass YSOs evolve more slowly and are also more abundant, increasing their chances of detection within the present MIR survey.

They are also however less luminous, causing them to fall below detection limits.

Finally, it is found that the YSOs have parametric solutions which extend over large ranges of value, despite the constraints upon $A_V$, $T_{EV}$ and $D$ cited above. This difficulty in pinning down narrow ranges of parameter leads to a smoothing of the histogramic trends.

So, it is apparent that a variety of problems may afflict the procedure described above, and make it difficult to define intrinsic trends in the YSO parameters. It is nevertheless clear that the graphs permit conservative limits to be placed upon these parameters. It would seem that the Class I sources are associated with a narrow range of temperatures $3.5 < log(T_{EFF}/K) < 3.65$, together with luminosities $L < 3 \; 10^2 \; L_\odot$ (a very small fraction has $L$ approaching $\sim 800 \; L_\odot$) and masses $M < 6.3 \; M_\odot$. The bimodal distribution of masses, should it prove to be real, may be an important indicator of how blast waves compress IS clouds, and lead to gravitational collapse and star formation. Finally, we note that envelope accretion rates are generally high ($1.6 \; 10^{-6} < (dM/dt)_{ENV}/M_\odot \; yr^{-1}) < 10^{-3}$) whilst disk masses appear to be modest ($\sim 10^{-2} \; M_\odot$). Stronger constraints will require deeper MIR surveys, however, and a more complete (and reliable) census of the Class I sources.

## 5. Conclusions

We have presented MIR imaging of four SNRs for which there appears to be prima-facie evidence for interaction with the ISM. We note for instance that certain shells appear to occupy voids in the MIR emitting environment, suggesting that the ISM has been impacted and removed, or that PAH emitting grains have been sputtered and vaporised. All of regions also show evidence for curved frontal structures, features which are likely to arise from the impact of SNR winds with ISM inhomogeneities and clouds. Finally, it has been noted that MIR ridge-like structures in W28 appear to be associated with regions of high velocity CO gas, and likely correspond to post-shock regimes in the north-easterly region of the shell; a regime for which OH 1720 MHz maser emission, enveloping HI shells, and shocked $H_2$ and ionic transitions testify to considerable levels of interaction with the ISM.

Almost all of the profiles imply flux ratios consistent with shock excitation of the v = 0→0 transitions of $H_2$, and the possible contribution of the CO fundamental band within the 4.5 μm channel, although we point out that care must be taken in the interpretation of these results. It is possible that stellar contamination is distorting these trends, and that much of the emission derives from PAH emission bands. Exceptions to these trends include a southerly region of interaction in W28, where the emission is clearly identified with PAH emission features, and a portion of a northerly front outside of the G355.6-00.0 shell, which may show evidence for [FeII] λ5.34 μm emission. Where PAH emission is enhanced, then this may imply that pre-existing grains have been shattered within local shocks, leading to larger abundances of smaller PAH emitting particles. The profiles also show evidence for a distinctive asymmetry, in which one side increases steeply, and the other declines more slowly. This latter, slower decline may arise as a result post-shock cooling and/or curvature of the shock.

Although several regions of interaction are known to be associated with enhanced radio surface brightnesses (viz. the NE region of W28), this is not, apparently, always necessarily the case. It is noted that radio surface brightnesses are often reduced in regions where the SNRs and ISM appear to be interacting (e.g. G001.0-00.1 and G355.9-02.5).

At least three of the sources also show evidence for triggered star formation, whereby interaction of the SNR and ISM leads to shocks, compression, and the formation of YSOs. Thus, we note that several YSOs are located at the periphery of G001.0-00.1, where evidence has previously been found for point source 1720 MHz maser emission. Similarly, several YSOs appear to be located on frontal structures close to the outer limits of the G355.9-02.5 outflow. Finally, we have noted that the spatial density of Class I YSOs within the boundaries of W28 is of order ~2-3 times greater than for adjacent regions of sky. This suggests that a good fraction of these stars may have been triggered by the SN event as well.

Given this likelihood, we have therefore analysed the properties of 11 Class I YSOs in the interaction region of the W28 shell. This is a source for which distances, extinctions, and the age of the shell are reasonably well defined. Using these constraints, in combination with the 2D radiative transfer modelling of Robitaille et al. (2006), and NIR-MIR SEDs deriving from GLIMPSE and 2MASS photometry, we are able to establish that the stellar masses, luminosities, radii and other parameters may be distributed over fairly large ranges, although trends are also affected by evolutionary and observational biases.

**Acknowledgements**

We would like to thank an anonymous referee for his interesting and perceptive report, and suggestions which led to several useful improvements to the analysis of this data. This work is based, in part, on observations made with the Spitzer Space Telescope, which is operated by the Jet Propulsion Laboratory, California Institute of Technology under a contract with NASA. Support for this work was provided by an award issued by JPL/Caltech. It also makes use of data products from the Two Micron All Sky Survey, which is a joint project of the University of Massachusetts and the Infrared Processing and Analysis Center/California Institute of Technology, funded by the National Aeronautics and Space Administration and the National Science Foundation. The 2MASS data was acquired using the NASA/IPAC Infrared Science Archive, which is operated by the Jet Propulsion Laboratory, California Institute of Technology, under contract with the National Aeronautics and Space Administration.

Guseinov O.H., Ankay A., Tagieva S.O., 2004, Serb. Astron., No. 169, 65

Hewitt J.W., Yusef-Zadeh F., Wardle M., 2008, ApJ, 683, 189

Hoffman I.M., Goss W.M., Brogan C.L., Claussen M.J., 2005, ApJ, 620, 257

Indebetouw R., et al., 2005, ApJS, 619, 931

Jones A.P., Tielens A.G.G.M., Hollenbach D.J., 1996, ApJ, 469, 740

Jones T.W., et al. 1998, PASP, 110, 125

Kassim N.E., 1989, ApJSS, 71, 799

Kundu M.R., 1970, ApJ, 162, 17

Kundu M.R., Velusamy T., 1972, A&A, 20, 237

LaRosa T.N., Kassim N.E., Lazio T.J.W., Hyman S.D., 2000, AJ, 119, 207

Liszt H., 1992, ApJS, 82, 495

Long K.S., Blair W.P., White R.L., Matsui Y., 1991, ApJ, 373, 567

Mehringer D.M., Goss W.M., Lis D.C., Palmer P., Menten K.M., 1998, ApJ, 493, 274

Melioli C., de Gouveia Dal Pino E. M., de La Reza R., Raga A., 2006, MNRAS, 373, 811

Mill J.M., O'Neil R., Price S.D., Romick G., Uy M., Gaposhkin E.M., J. Spacecraft & Rockets, 31, 900

Milne D.K., 1979, Austr. J. Phys., 32, 83

Milne D.K., Wilson, T.L., A&A, 10, 220

# Figure Captions

**Figure 1**

The region of SNR G001.0-00.1, where we illustrate emission at 8.0 $\mu$m (red); the 90 cm VLA radio mapping of LaRosa et al (2000) (green); the 1720 MHz contours of OH thermal emission from Yusef-Zadeh (1999) (white); and the positions of Class I YSOs (blue circles). The region to the upper left of the panel corresponds to an HII region. Note how the SNR appears to be excavating a cavity in the 8.0 $\mu$m emitting material, and how many of the YSOs are located close to the periphery of the shell.

**Figure 2**

Radio and MIR image of a globule to the east of G001.0-0.1. The upper panel illustrates how the convex front of the globule is immersed within the SNR, and reveals a tail of ram-pressure stripped material to the lower left-hand side. The infrared image is composed of 5.8 $\mu$m and 8.0 $\mu$m results, represented respectively as yellow and red, whilst the inserted panel shows the globule at higher degrees of saturation, and reveals fainter components of wind-swept material. The profiles through this structure (lower panel) show that emission is strongest in a narrow rim around the source, presumably arising from shock interaction with the SNR. The inserted figure show The directions and widths of the profiles are indicated as a white bar within the upper panel, and in the insert of the lower panel, whilst parameters ($\Delta 3.6$, $\Delta 4.5$, $\Delta 5.8$, $\Delta 8.0$) are given by (6.729, 5.688, 23.704, 71.619) MJy sr$^{-1}$.

**Figure 3**

The positions of profile flux ratios for our sample of SNRs within the colour-colour diagnostic diagram of Reach et al. (2006), where the zones of synchrotron, $H_2$, ionic and PAH emission are separately indicated. Most of the sources appear to be located in the region of molecular emission, although these values may also be affected by stellar contamination. Where this is the case, then the sources would move towards the area defining PAH emission.

**Figure 4**

As for Fig. 1, but for the case of G355.9-02.5, where we have represented 8.0 μm emission as red, and the 843 MHz emission of Gray (1994b) by green (upper panel). The yellow circles indicate the positions of Class I YSOs. Note how the SNR shell appears to interact with the ISM to the lower left-hand side, leading to a band of radio emission parallel to the ridge of MIR emitting material. There is also evidence for various parabolic frontal structures in the lower part of the image. These latter structures are more clearly illustrated in the lower panel of the figure, where we have indicated the directions of the profiles in Fig. 5. Note how three of the five YSOs (green circles) appear to lie close to frontal structures.

**Figure 5**

MIR profiles for two of the frontal structures in G355.9-02.6. The behaviours of these fronts appears to be markedly different. In the upper panel, the emission in structure A peaks at the presumed position of the front (RP = 0 arcsec), and thereafter declines rapidly to negative RPs. There is very little emission at 3.6 or 4.5 μm, although we note evidence for a possible Class I YSO at RP = 0 arcsec. By contrast, the emission at frontal structure B increases rapidly at RP = 0 arcsec, and then stays more-or-less constant to negative RPs. The parameters ($\Delta 3.6$, $\Delta 4.5$, $\Delta 5.8$, $\Delta 8.0$) for the upper panel are (2.52, 1.431, 4.493, 13.443) MJy sr$^{-1}$, and those for the lower panel are (2.954, 1.927, 4.105, 9,271) MJy sr$^{-1}$.

**Figure 6**

Superimposition of the 843 MHz radio image due to Gray (1994a) (indicated as green) upon 5.8 (yellow) and 8.0 μm (red) mapping of G355.6-00.0 (upper panel; see the text for details). The SNR shell is represented by the roughly circular structure in the lower central portions of the image, whilst emission to the lower right corresponds to an HII region. Note the evidence for a band of higher extinction material along the latitudes b ~ 0.04-0.06°; diffuse ionised emission above the SNR; and a narrow frontal structure close to l = 355° 41' 42", $b$ = +00° 02' 50" (coloured yellow). The blue circles indicate the positions of Class I YSOs. It is suggested that the extinction band and putative ionisation or shock front may represent regions of interaction with the SNR. An expansion of the shock/I-front region is also illustrated in the lower panel, where we indicate the directions and widths of the profiles in Fig. 7.

**Figure 7**

Profiles through the frontal structure close to G355.6-00.0, where the widths and directions of the profiles are indicated in the inserted panels. Note the strong increase in emission to longer MIR wavelengths, the relative narrowness of the frontal structure, and the evidence for a Class I YSO in profile A (at RP = 0 arcsec). The vertical axis corresponds to the relative variation in surface brightness. To retrieve absolute values for this parameter, one should add the quantities ($\Delta 3.6$, $\Delta 4.5$, $\Delta 5.8$, $\Delta 8.0$) = (5.246, 3.805, 23.102, 69.273) MJy sr$^{-1}$ to the profiles in the upper panel, and ($\Delta 3.6$, $\Delta 4.5$, $\Delta 5.8$, $\Delta 8.0$) = (4.175, 3.608, 22.205, 69.277) MJy sr$^{-1}$ to profiles in the lower panel.

**Figure 8**

Superposition of the 1415 MHz radio observations of Dubner et al. (2000) (green), the Rosat X-ray observations of Rho & Borkowski (2002) (blue), and the present *Spitzer* 5.8 (yellow) and 8.0 $\mu$m (red) observations in the region of W28. The positions of Class I YSOs are indicated using blue circles, whilst the approximate outer limits of the SNR shell are delineated with a white contour. Note that a thin frontal structure appears to extend down the left-hand side of the image, and curves about the bottom. Whilst the relation between this structure to the SN event is far from well established, it is possible that it was generated through interaction between pre-cursor winds and the ISM.

**Figure 9**

Two regions of interaction within the W28 SNR shell. In the upper panel, the red-yellow emission corresponds to fluxes at 5.8 and 8.0 $\mu$m. By contrast, the 327 MHz MHz mapping of Frail et al. (1993) is indicated using white contours, whilst the region of high velocity CO post shock gas (Arikawa et al. 1999; see also Hoffman et al. 2005) is defined by the green contours. The small white circles indicate the locations of Class I YSOs, whilst blue crosses indicate the sources used in our study of YSO parameters (Sect. 4). The lower panel, by contrast, indicates an apparent region of interaction between an ISM globule and the SNR. In both cases, the white/dark bars indicate the directions and widths of the profiles in Figs. 10 and 11.

**Figure 10**

Profiles through two bars of MIR emission associated with high velocity CO gas in W 28, where the widths and orientations of the profiles are indicated in the inserted panels. The parameters ($\Delta 3.6$, $\Delta 4.5$, $\Delta 5.8$, $\Delta 8.0$) for the upper panel are (5.389, 3.47, 31.839, 104.695) MJy sr$^{-1}$, and those for the lower panel are (7.77, 5.09, 31.0, 100.2) MJy sr$^{-1}$.

**Figure 11**

As for Fig. 10, but for a frontal structure at lower Galactic latitudes. Note the steep rise in emission between RPs of 80 and 90 arcsec, and the somewhat gentler fall-off to lower RPs. Absolute values of surface brightness can be obtained by applying the corrections ($\Delta 3.6$, $\Delta 4.5$, $\Delta 5.8$, $\Delta 8.0$) = (5.7, 1.9, 25.08, 81.75) MJy sr$^{-1}$.

**Figure 12**

The positions of YSOs in W28 as plotted in the MIR colour plane, where the small blue symbols correspond to Class I sources located within the primary shell of W28 (illustrated in Fig. 8), and the large red symbols correspond to triggered YSOs in the NE region of interaction, illustrated in the upper panel of Fig. 9 (blue crosses). We finally indicate the positions of various categories of YSO (based on the analysis of Allen et al. (2004)), the mean error in the colour estimates, and the reddening vector for $A_V$ = 20 mag.

**Figure 13**

The results of fitting 2-D radiative transfer modelling to the SED of a YSO at $l$ = 6.535°, $b$ = -0.1009°, where a variety of parameters are represented against the age of the source. Individual bullets indicate the results of best fit models for which $\chi^2 - \chi_{BEST}^2 < 3$, whilst the grey shading delineates the ranges of model parameters which were explored. Where the YSO was formed as a result of the W28 event, then the age can be constrained to be less than ~3.5 10$^4$ yrs (see text for details). This restriction permits us to limit the physical characteristics of the source. Thus, the temperature of the star is more likely to be ~3500 K, rather than the slightly higher values applying for ages of several millions of years.

**Figure 14**

The normalized distributions of physical parameters for 12 YSOs within the interaction region of W28, illustrated in the upper panel of Fig. 9. If the stars were formed as a result of triggering by W28, then it is clear that the ranges of physical parameter are relatively broad. There is no particular trend in favour of lower luminosities and masses, a peculiarity which probably arises due to biases in the detection of the sources, and the broad ranges of the model solutions. However, there may be evidence for a bilobal distribution of masses which could, if real, have a bearing upon shock induced star-formation processes.

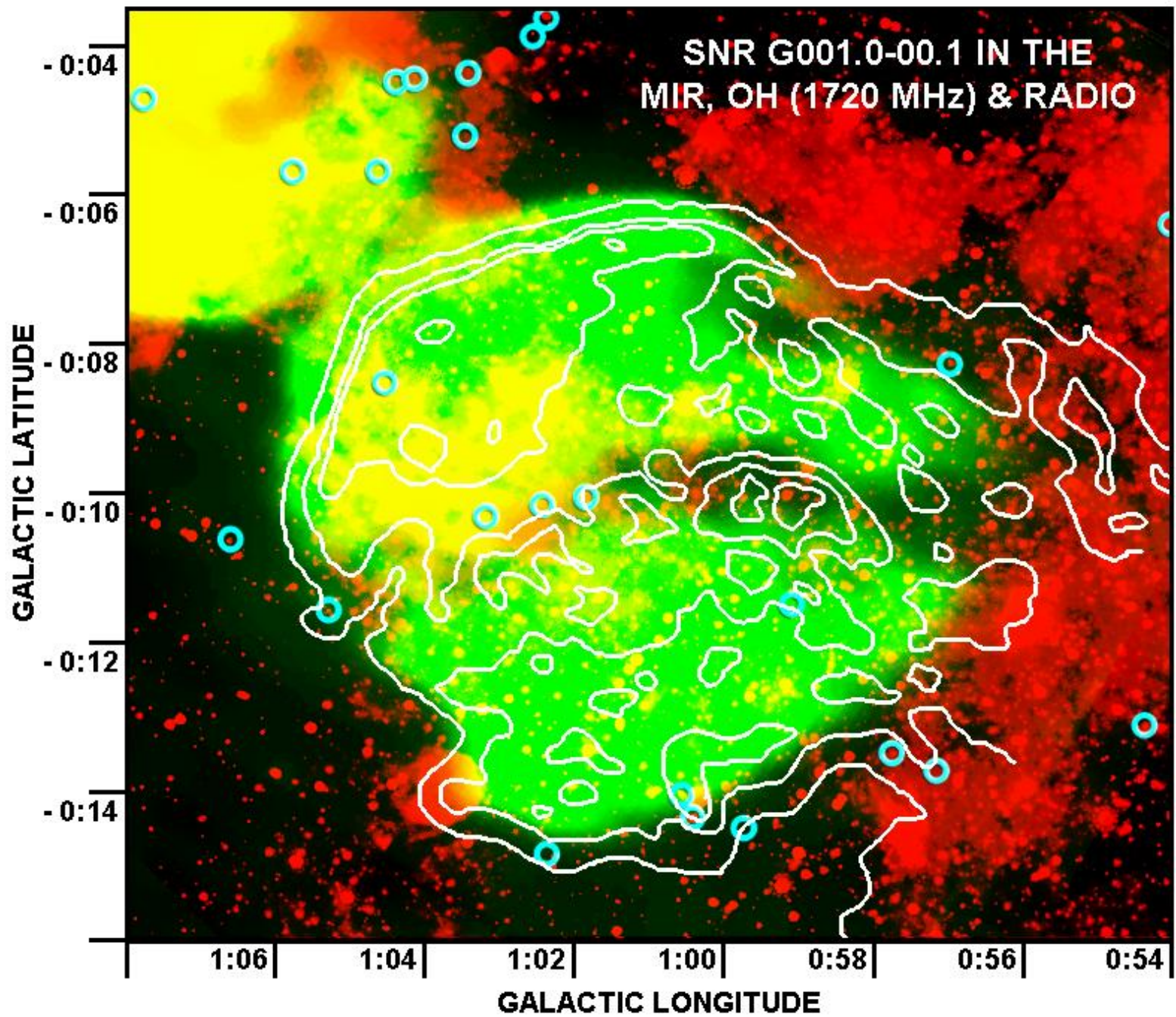

FIGURE 1

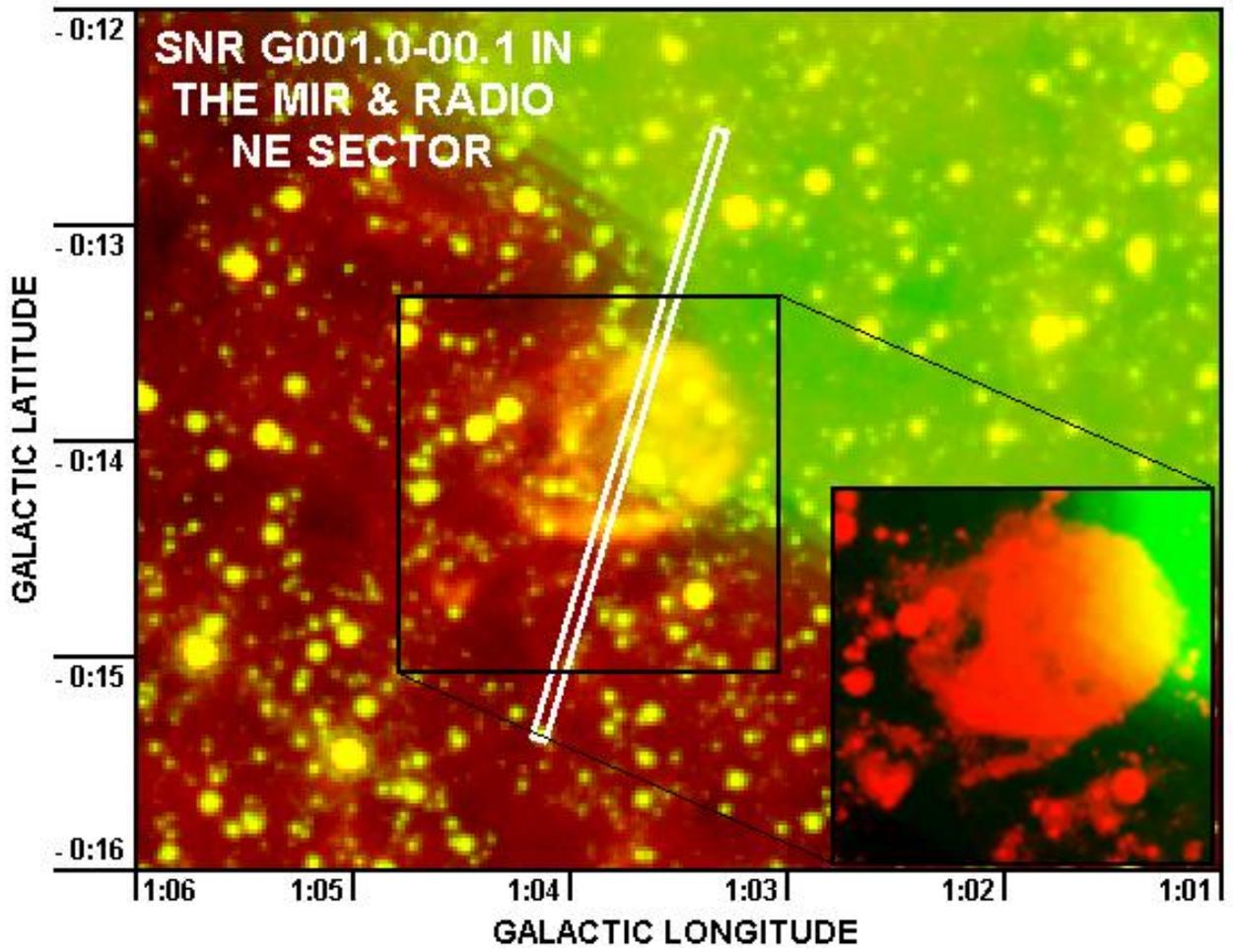
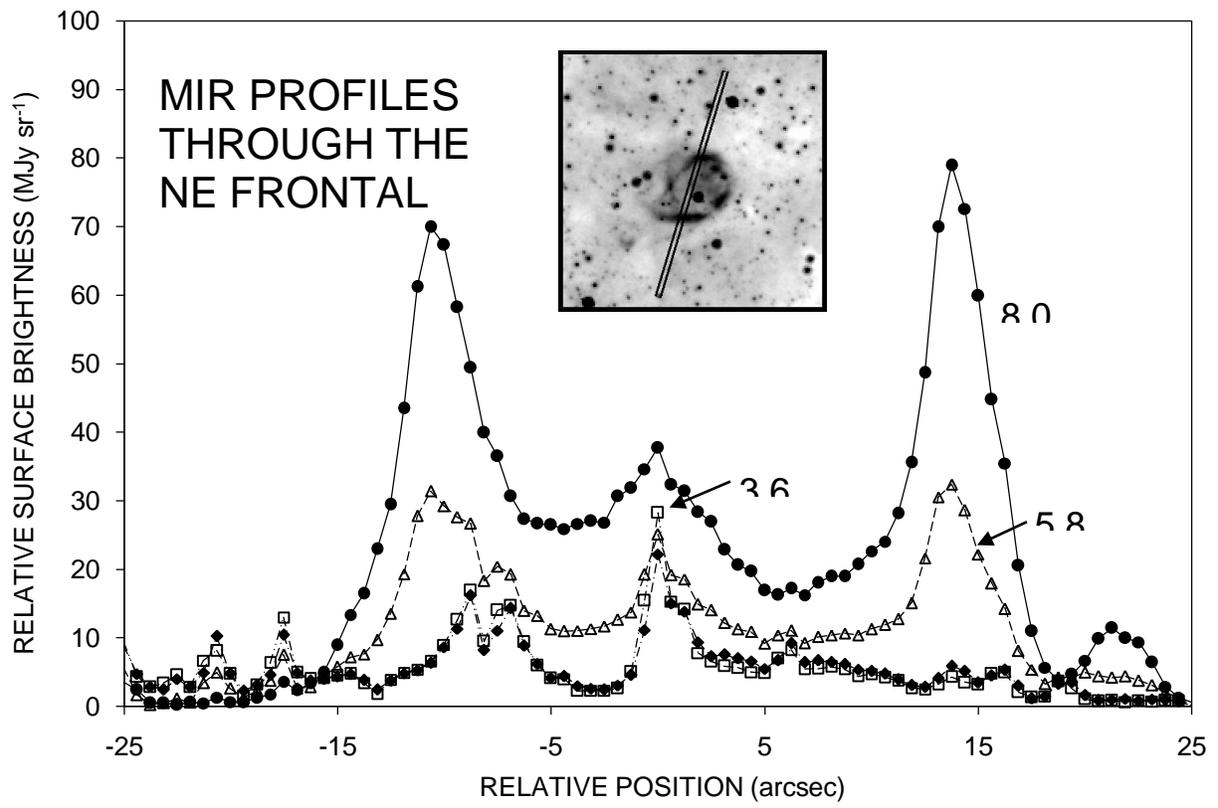

FIGURE 2

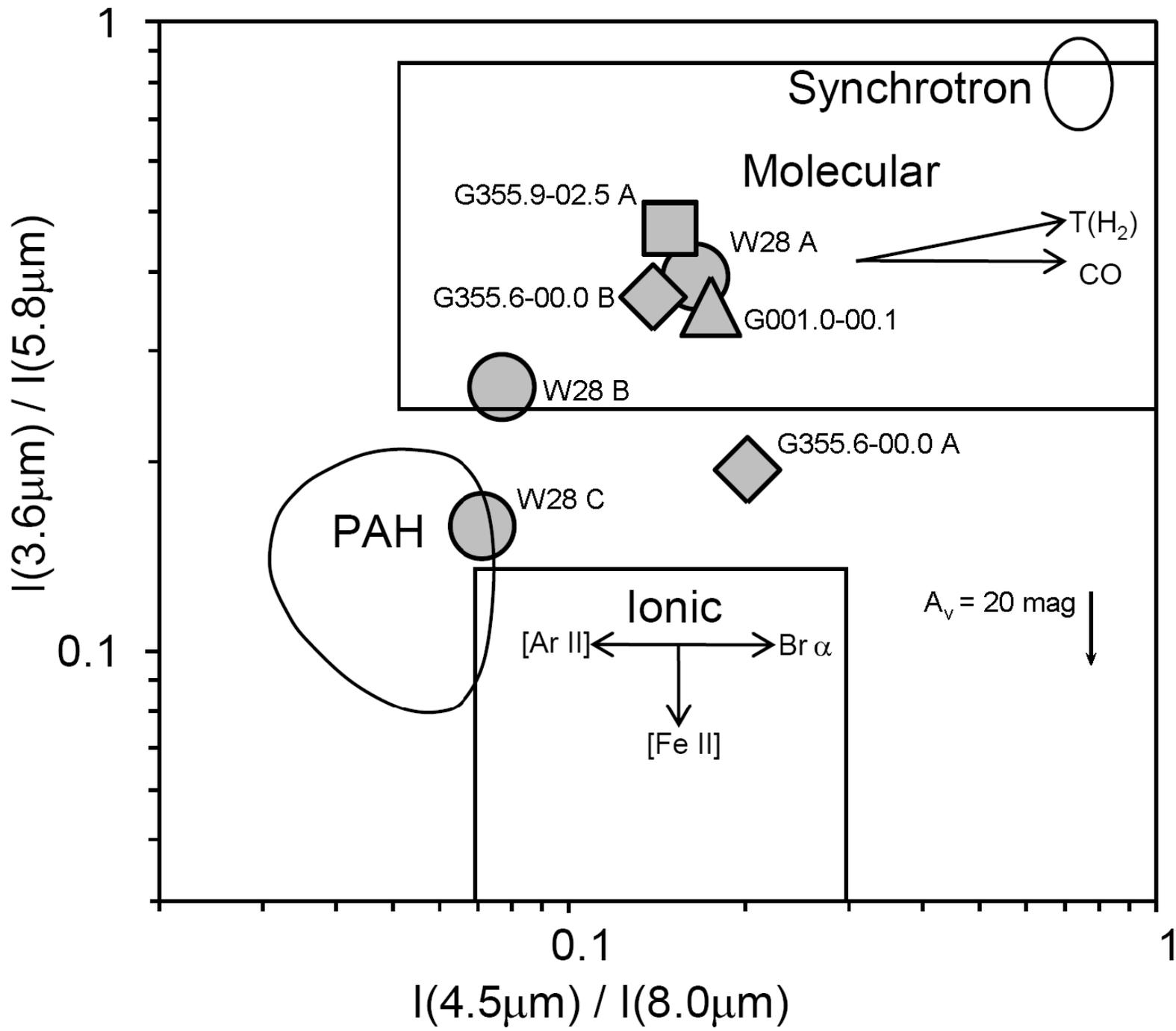

FIGURE 3

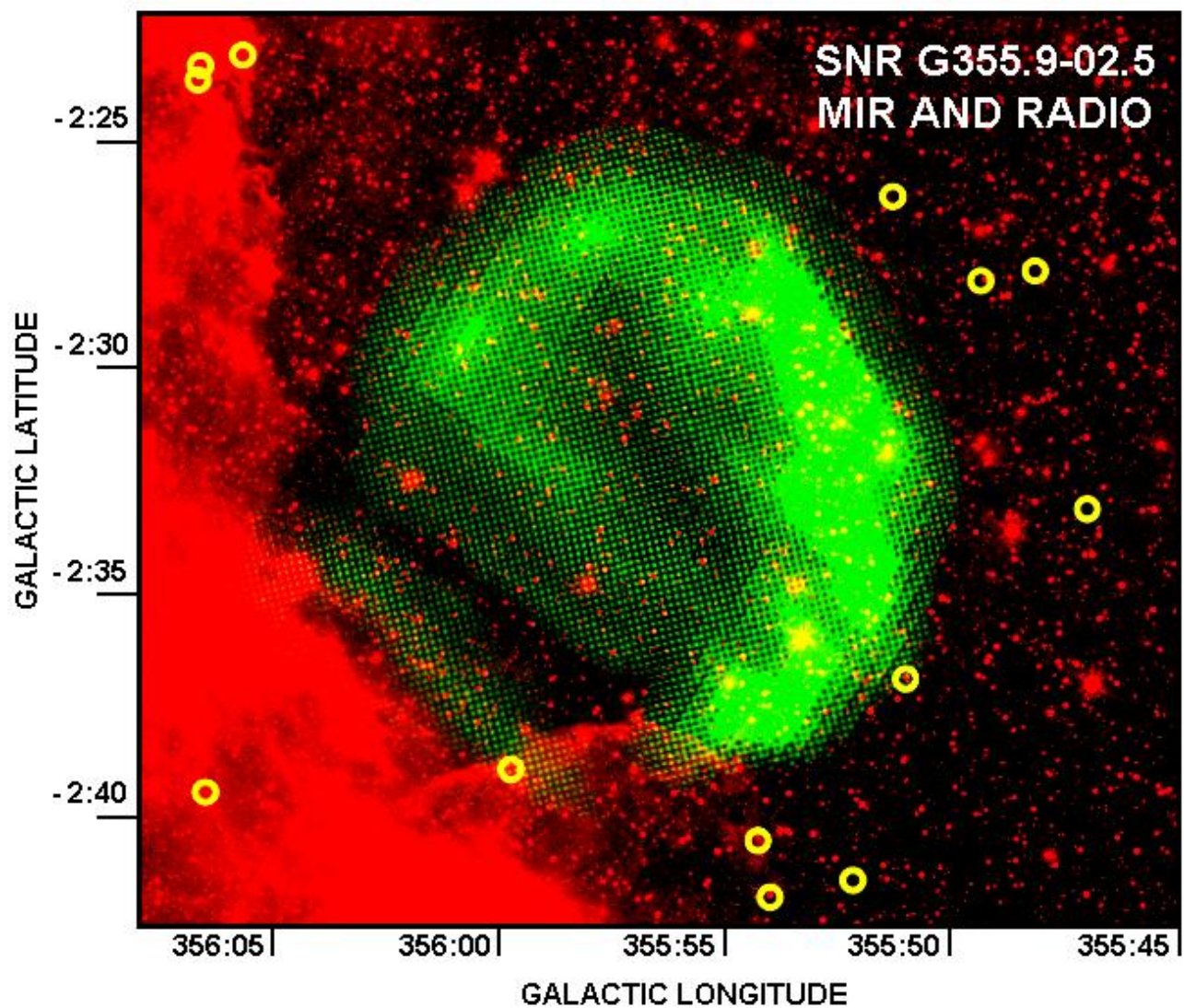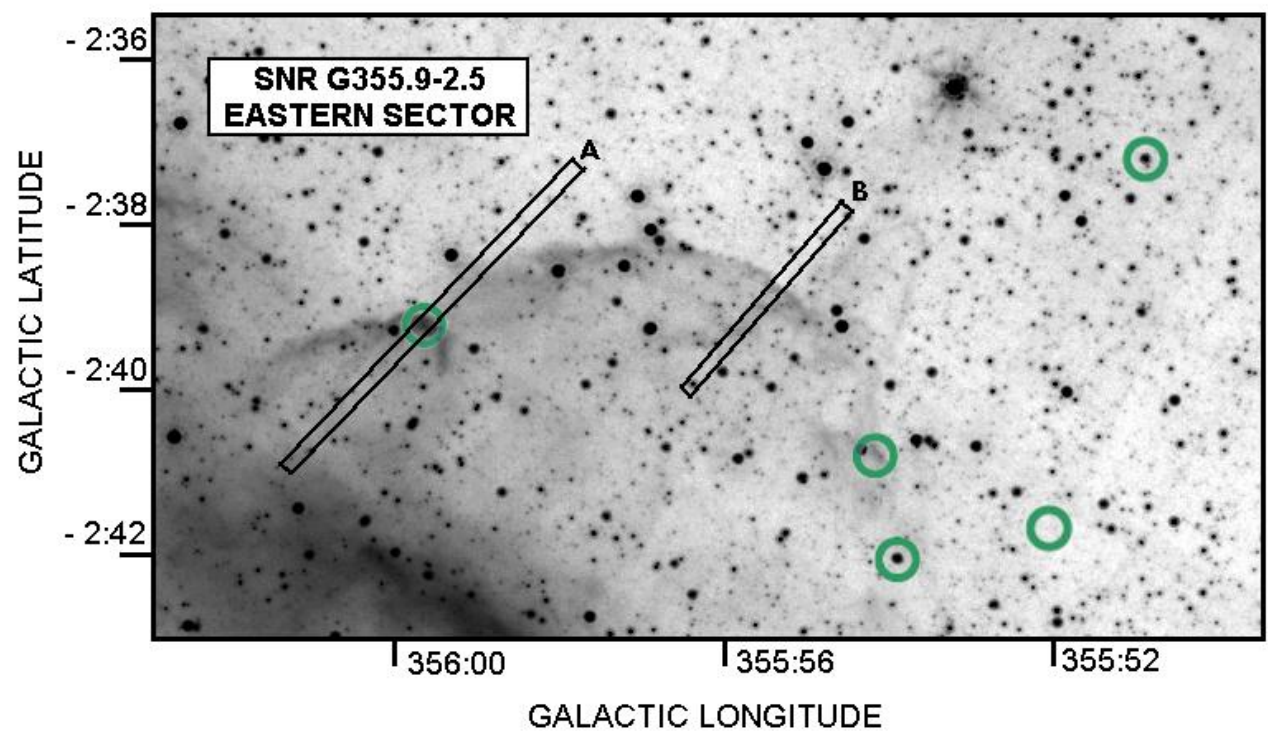

FIGURE 4

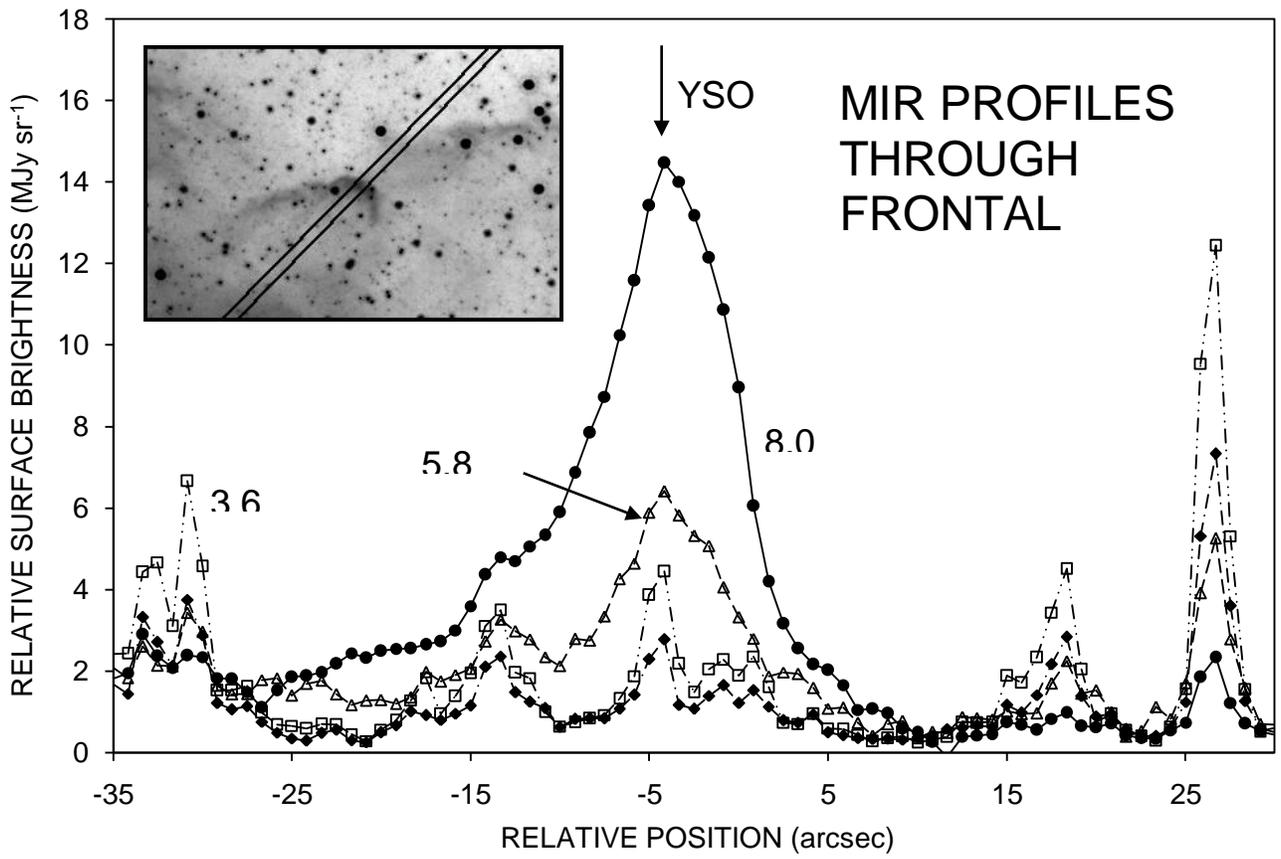

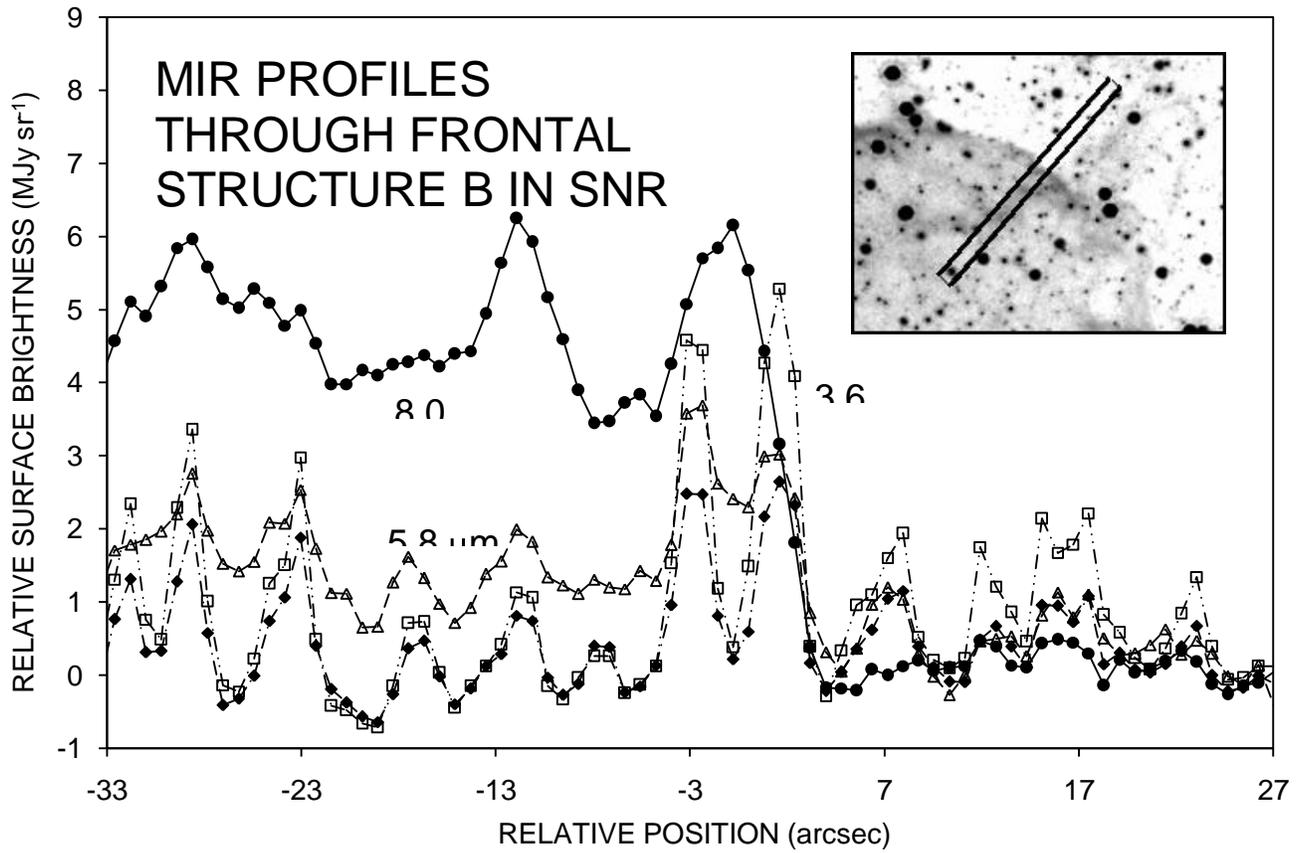

FIGURE 5

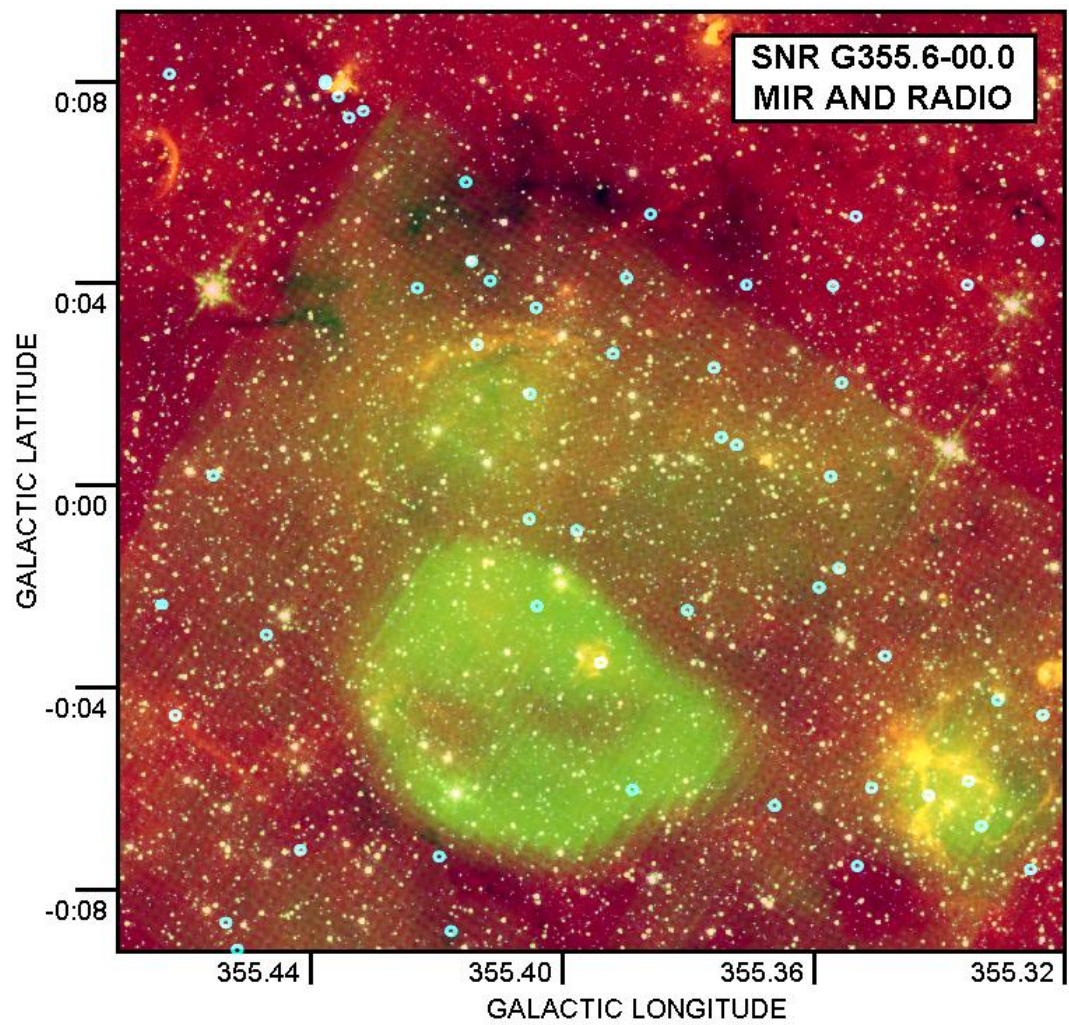
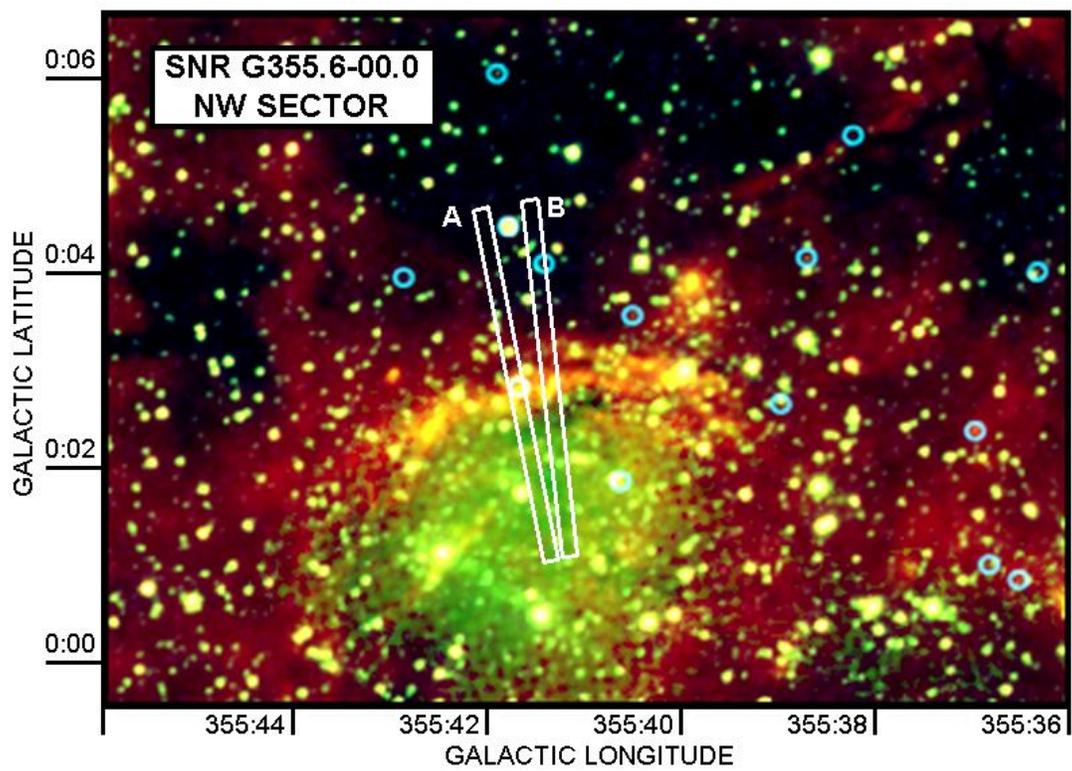

FIGURE 6

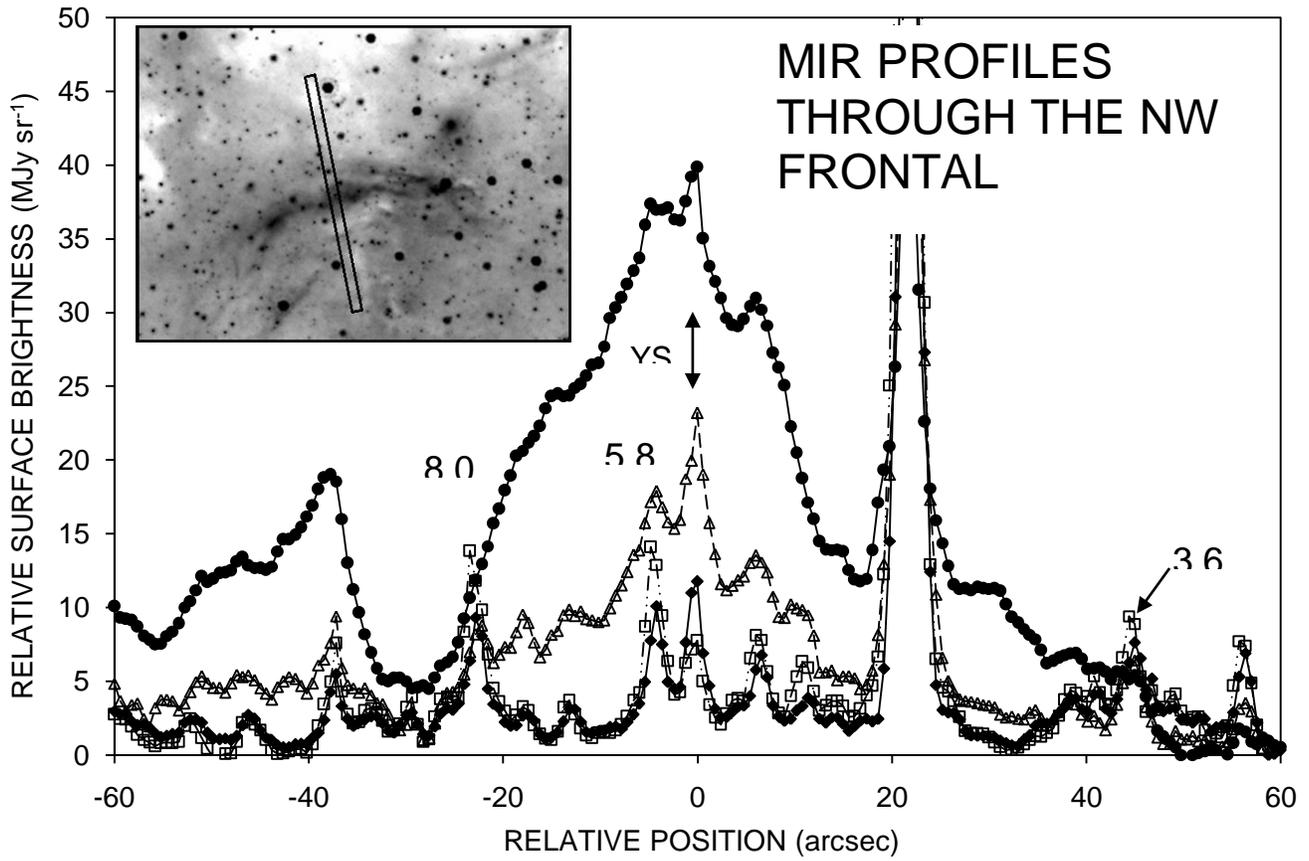
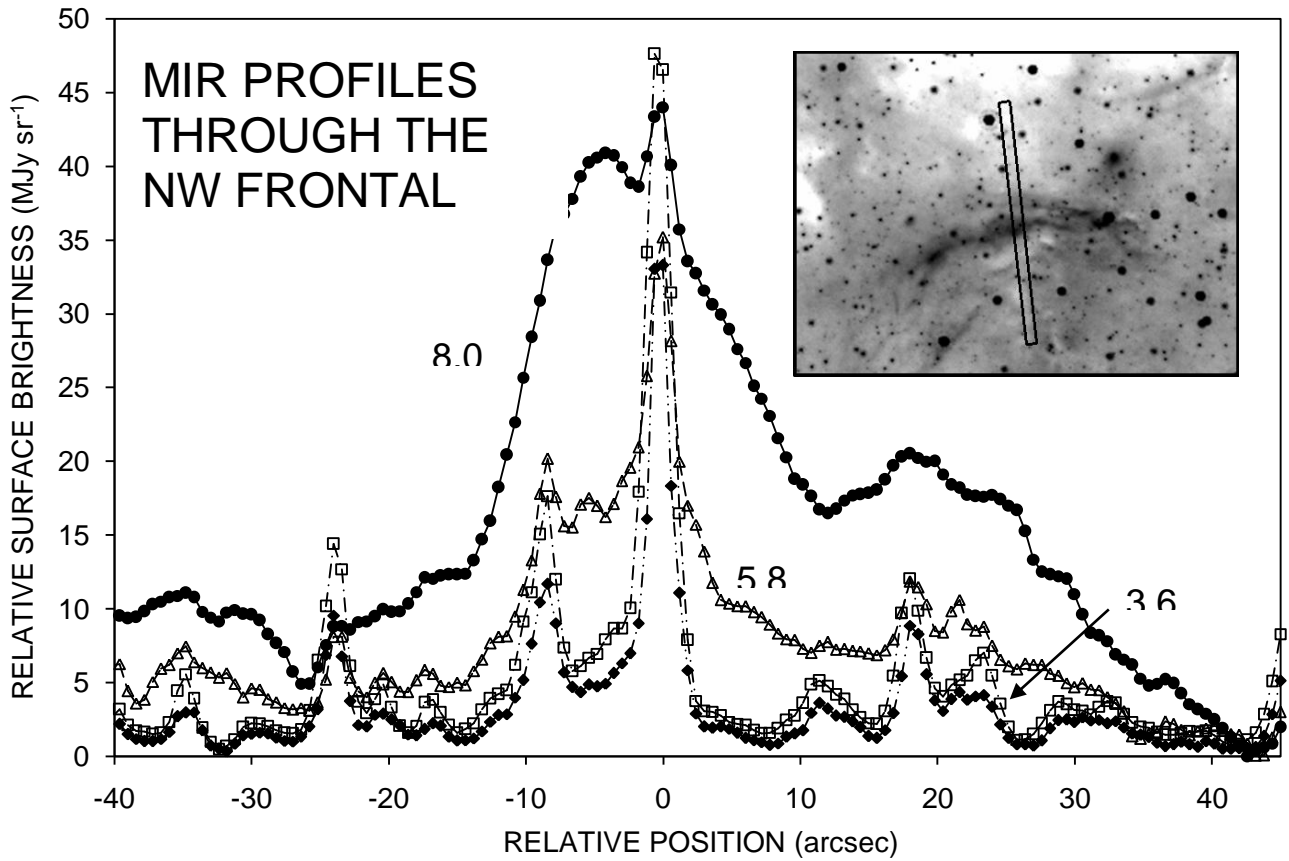

FIGURE 7

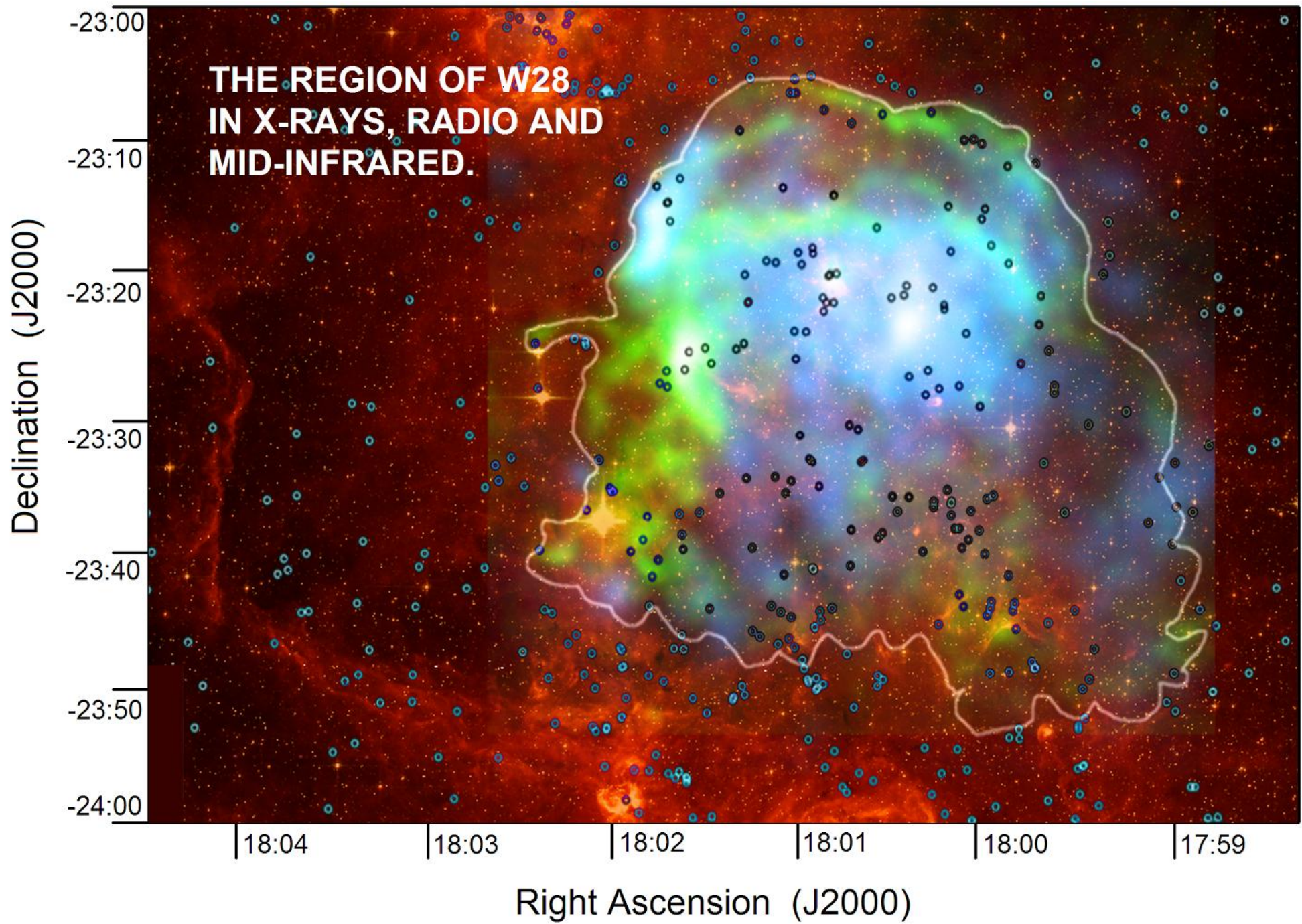

FIGURE 8

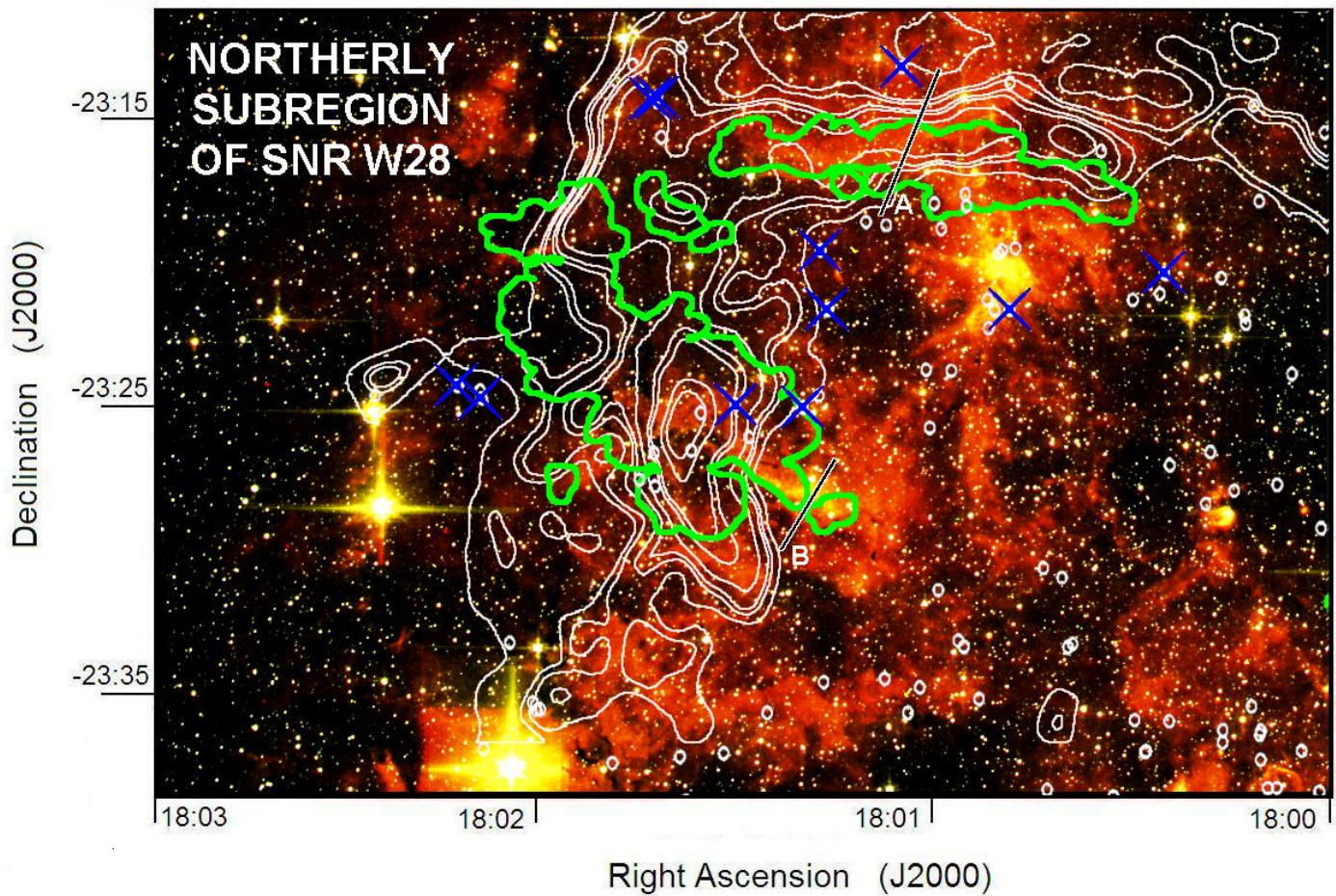
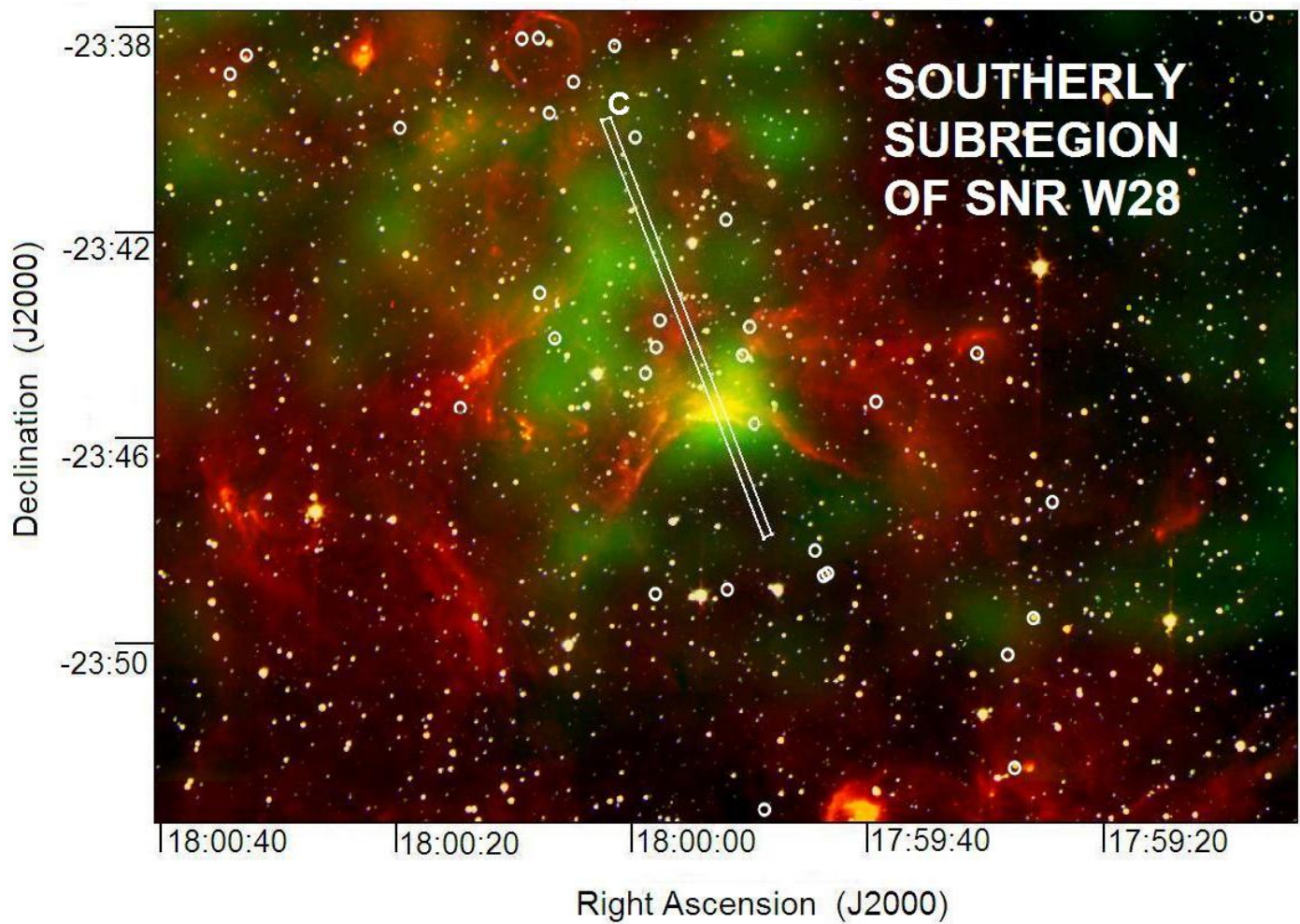

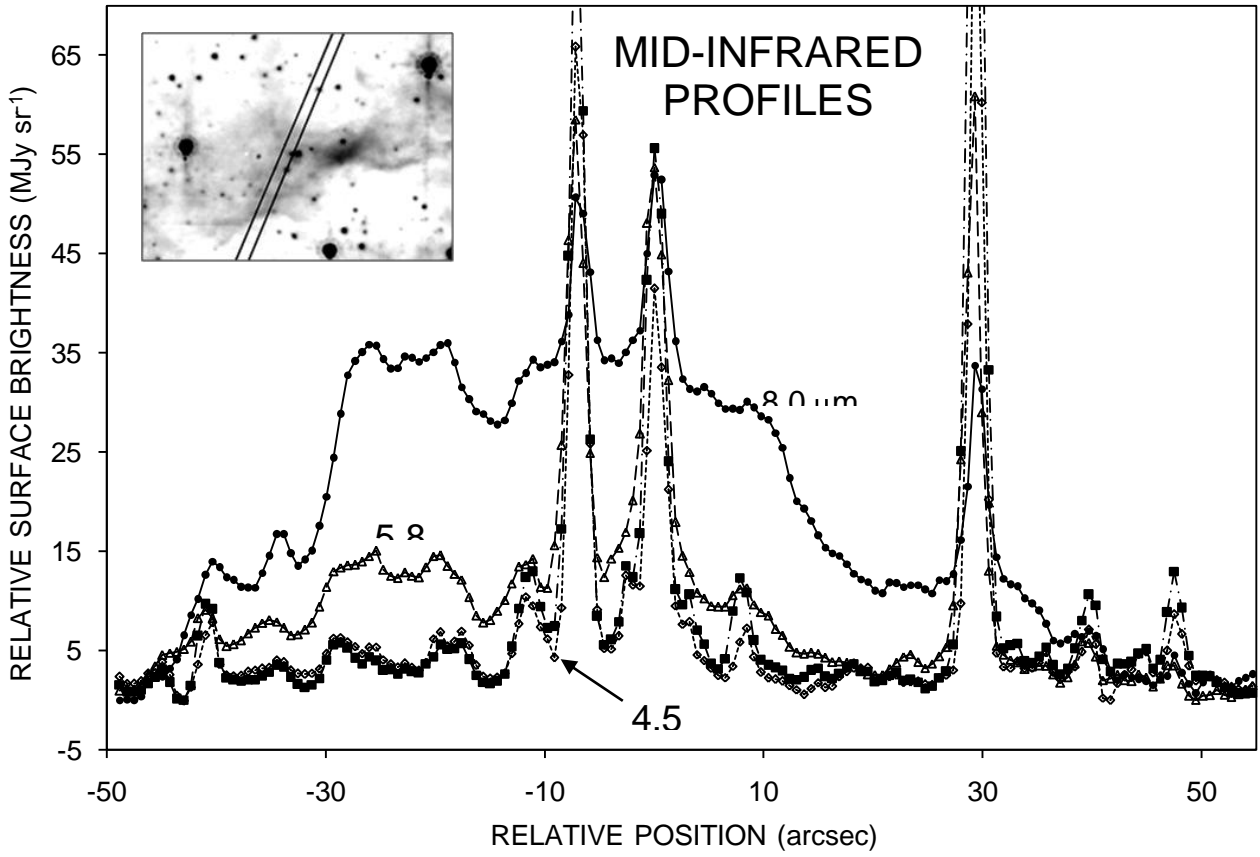

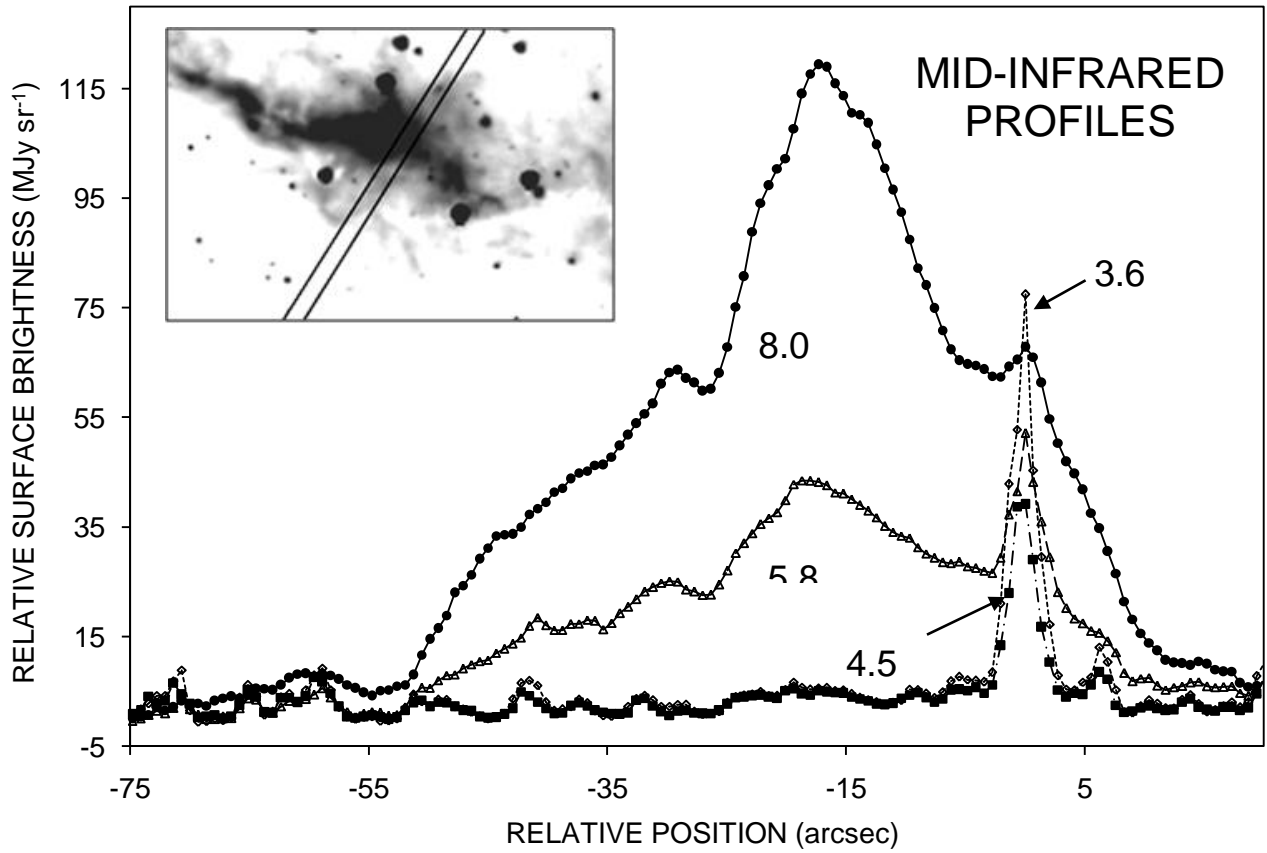

FIGURE 10

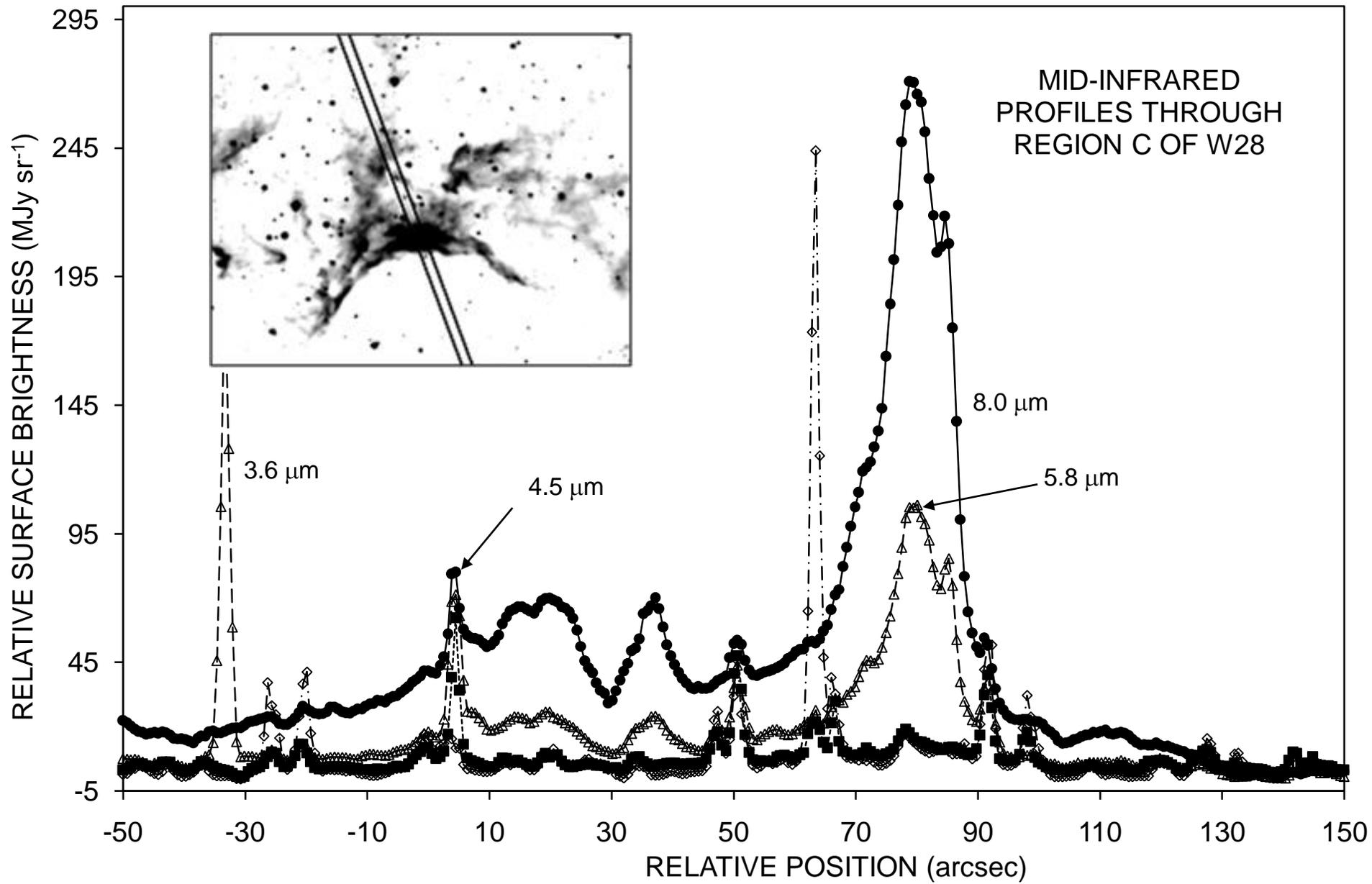

FIGURE 11

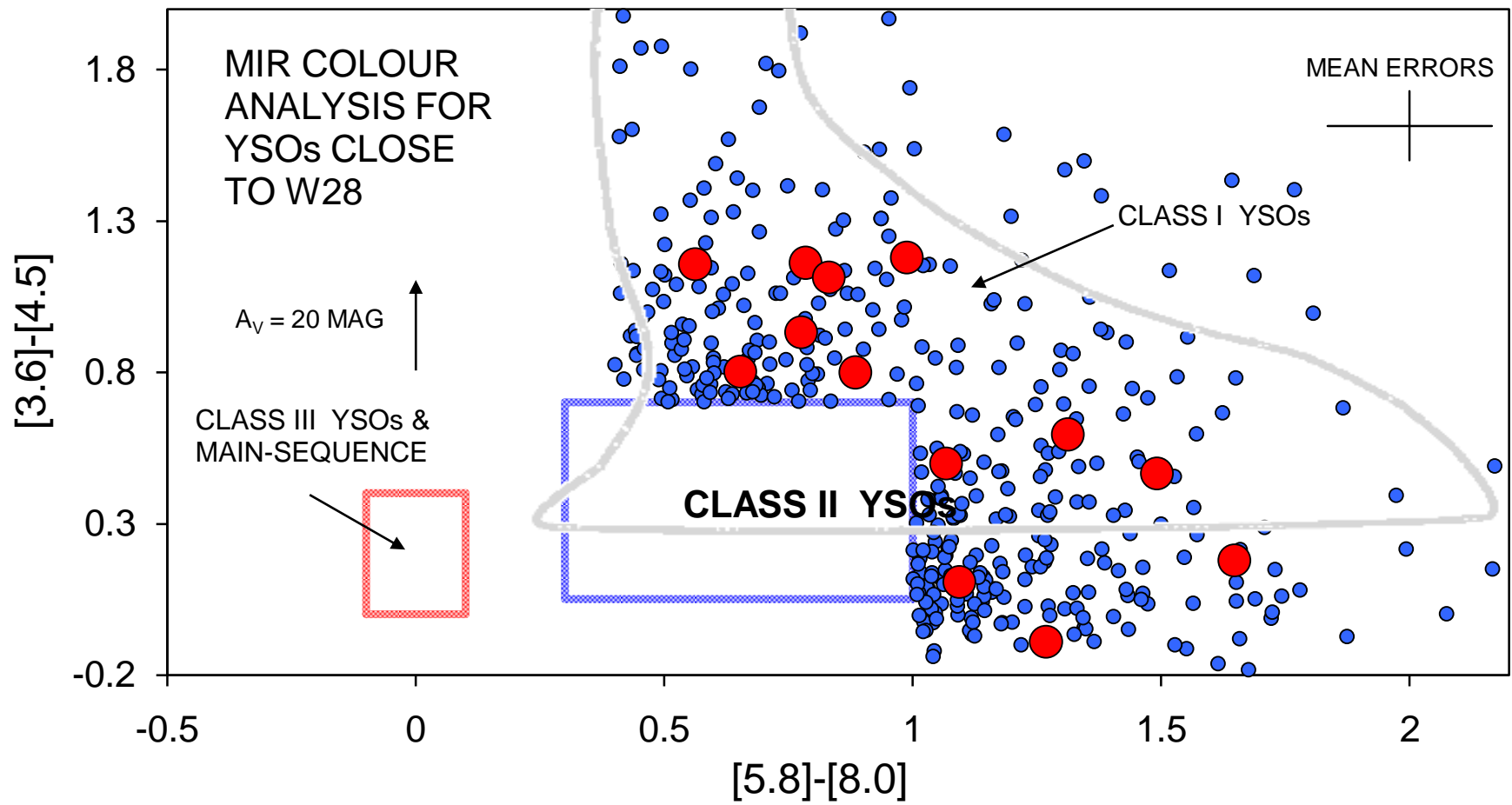

FIGURE 12

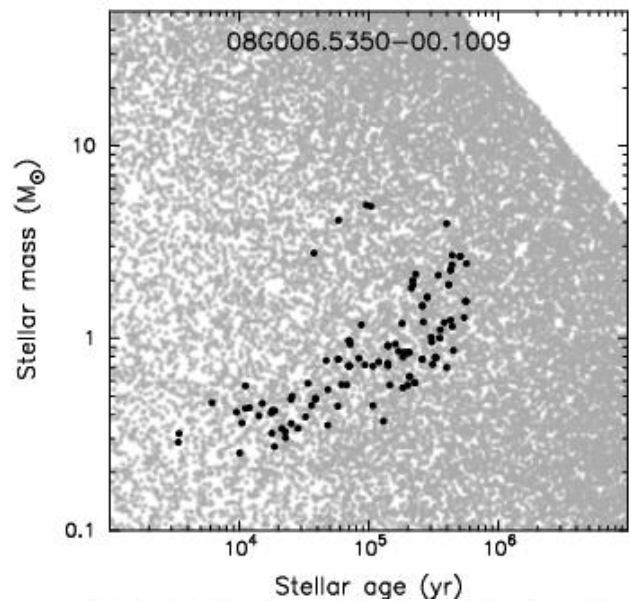
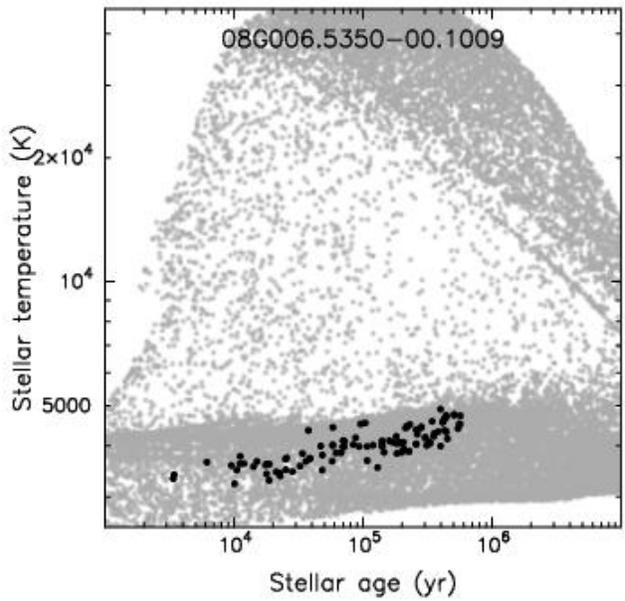
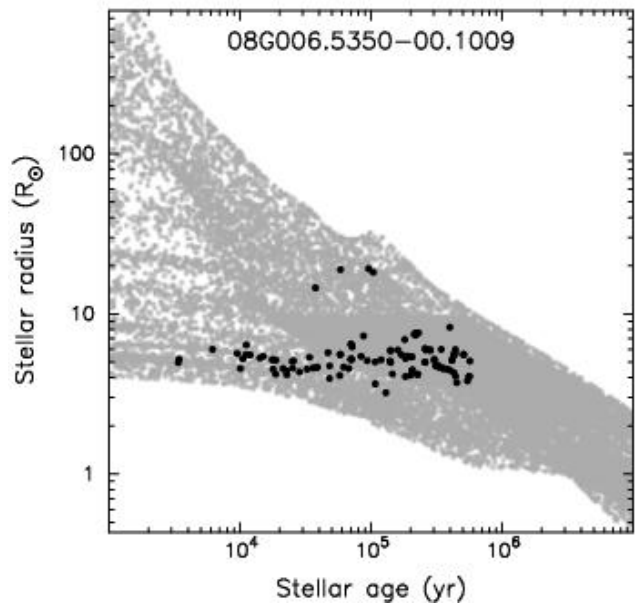
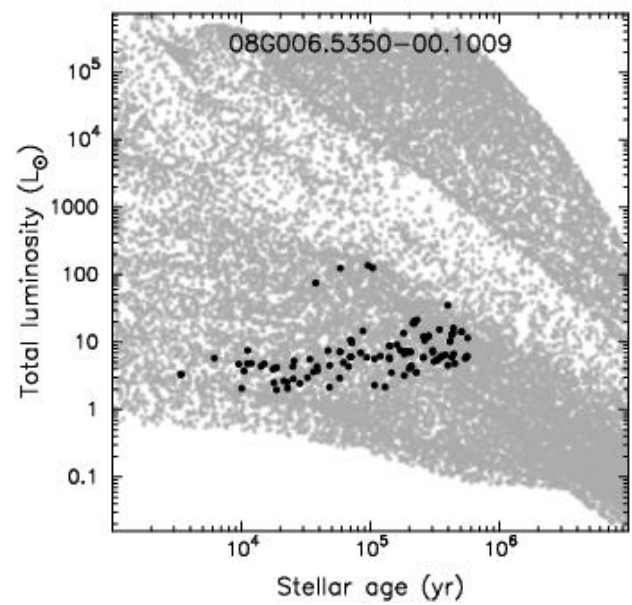
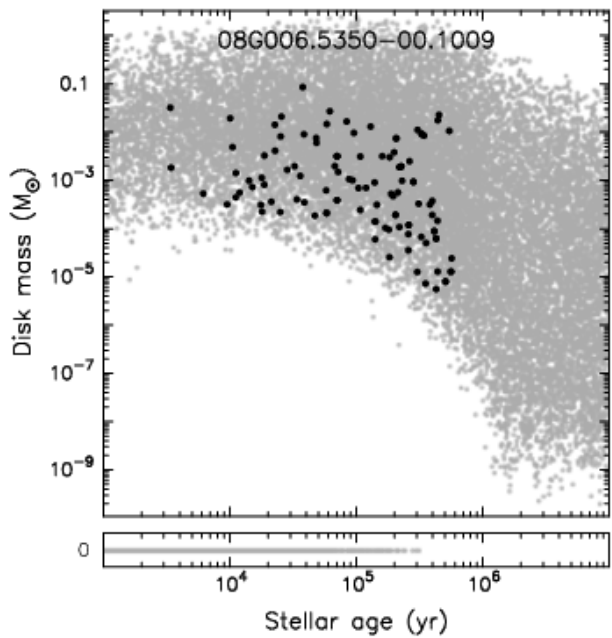
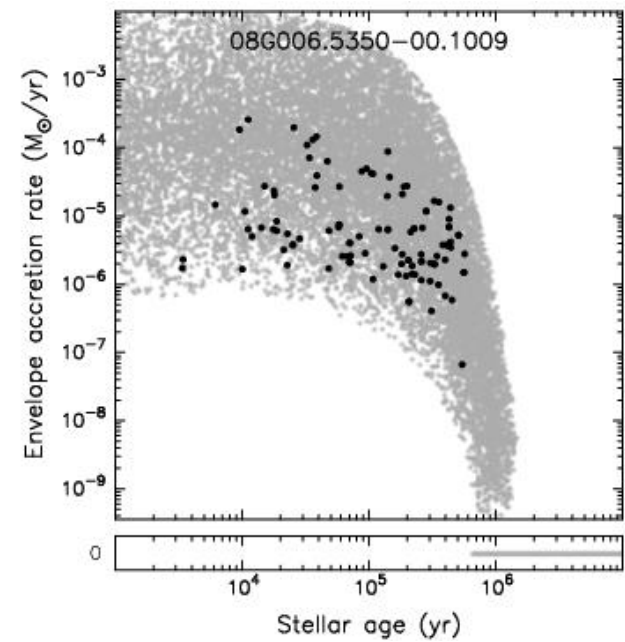

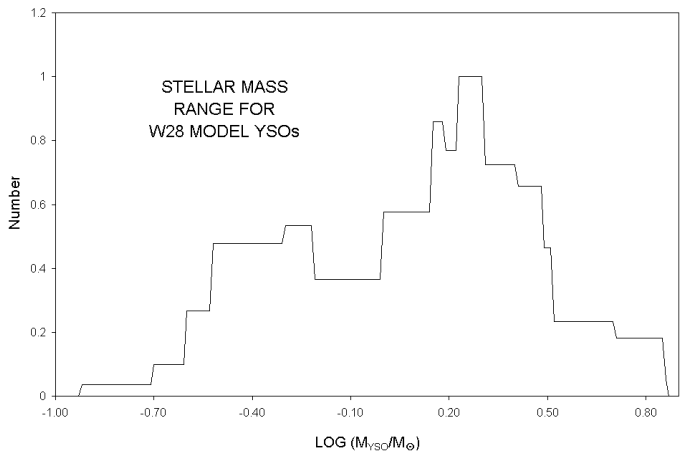
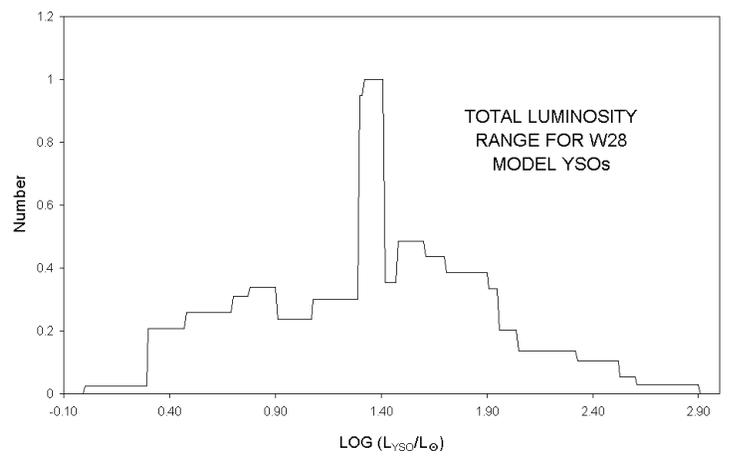
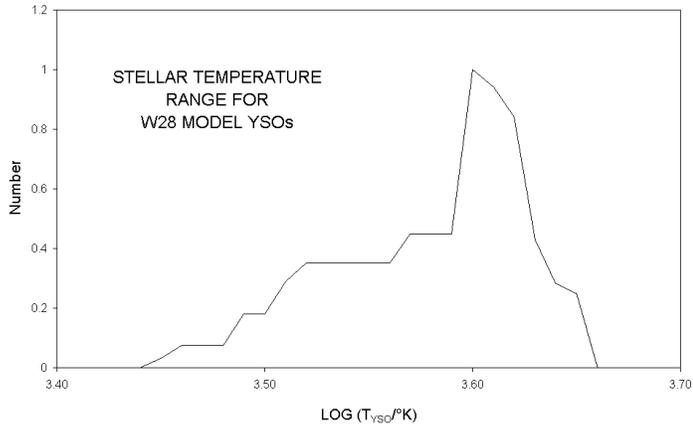
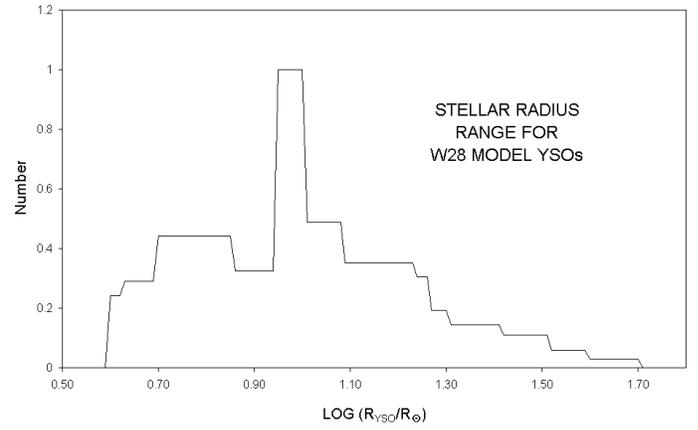
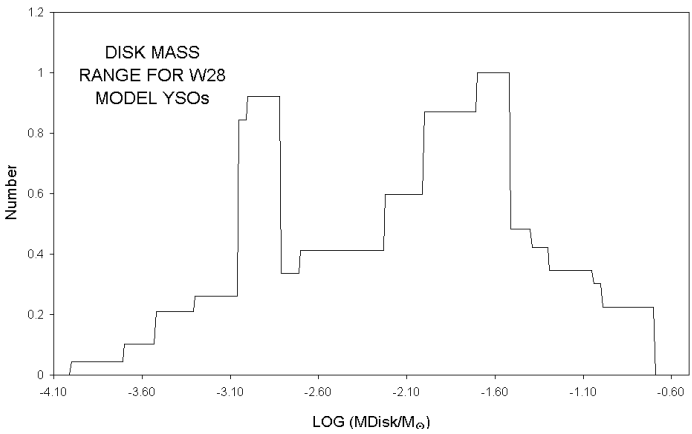
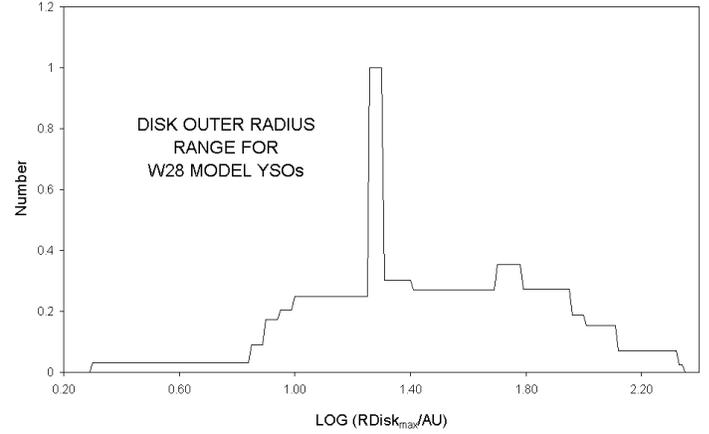
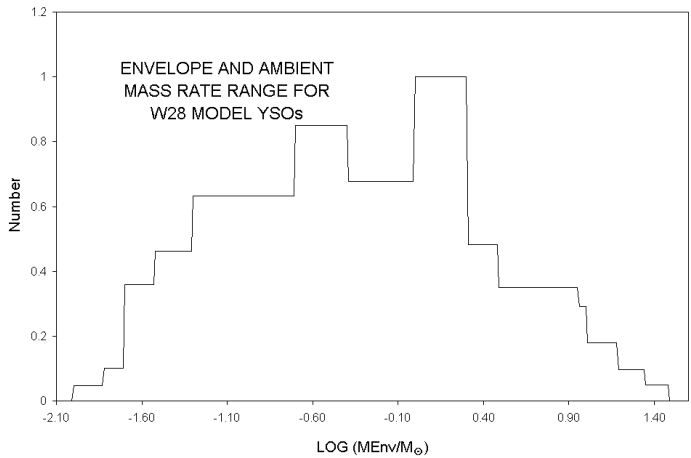
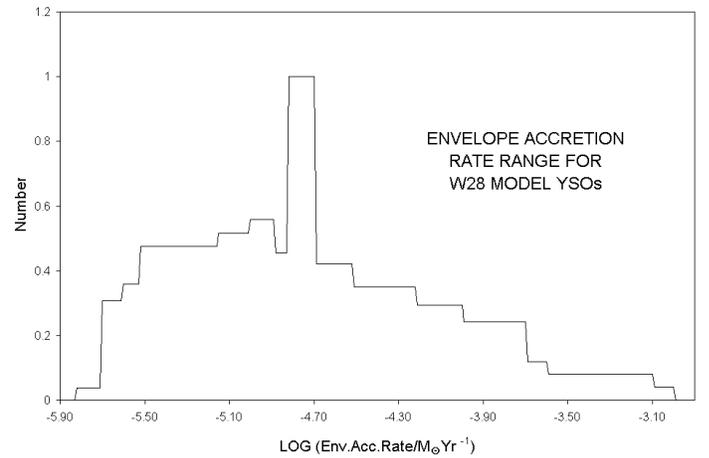